\documentclass[oneside,journal]{IEEEtran}
\usepackage{amsmath,amsfonts}
\usepackage{algorithmic}
\usepackage{algorithm}
\usepackage{array}
\usepackage{textcomp}
\usepackage{stfloats}
\usepackage{url}
\usepackage{verbatim}
\usepackage{graphicx}
\usepackage{cite}
\usepackage{upgreek}
\usepackage{amssymb} 
\usepackage{xcolor}
\usepackage{balance}
\usepackage{hyperref}
\ifCLASSOPTIONcompsoc
	\usepackage[caption=false, font=normalsize, labelfont=sf, textfont=sf]{subfig}
\else
	\usepackage[caption=false, font=footnotesize]{subfig}
\fi
\newtheorem{remark}{Remark}
\newtheorem{theorem}{Theorem}
\newtheorem{lemma}{Lemma}

\newtheorem{corollary}{Corollary}
\renewcommand{\arraystretch}{1.25}
\begin{document}

\title{Coordinated  Beamforming   for Networked Integrated   Communication and Multi-TMT Localization}


\author{Meidong~Xia,~\IEEEmembership{Student Member,~IEEE,}
		Zhenyao~He,~\IEEEmembership{Student Member,~IEEE,}
		Wei~Xu,~\IEEEmembership{Fellow,~IEEE},
		Yongming~Huang,~\IEEEmembership{Fellow,~IEEE},  
		Derrick~Wing~Kwan~Ng,~\IEEEmembership{Fellow,~IEEE},
		and Naofal~Al-Dhahir,~\IEEEmembership{Fellow,~IEEE}
		
		\thanks{
		The work of Naofal Al-Dhahir was supported by Erik Jonsson Distinguished Professorship at UT-Dallas.
		Part of this work has been accepted for presentation at the 2025 IEEE
		International Conference on Communications (ICC)~\cite{xiaMultiCellCoordinatedBeamforming2025}. 
		\textit{(Corresponding author: Wei Xu.)}}
		\thanks{Meidong  Xia, Zhenyao He, Wei Xu, and Yongming Huang 
		are with the National Mobile Communications Research Laboratory, 
		Southeast University, Nanjing 211189, China, 
		and also with the Purple Mountain Laboratories, Nanjing 211111, China (e-mail: \{meidong.xia, hezhenyao, wxu, huangym\}@seu.edu.cn).}
		\thanks{Derrick Wing Kwan Ng is with the School of Electrical Engineering and Telecommunications, 
		University of New South Wales, Sydney, NSW 2052, Australia (e-mail: w.k.ng@unsw.edu.au).}
		\thanks{Naofal Al-Dhahir is with the Department of Electrical and Computer
		Engineering, the University of Texas at Dallas, Richardson, TX 75080, USA (e-mail: aldhahir@utdallas.edu).}
		}

\maketitle

\begin{abstract}
Networked integrated sensing and communication (ISAC) has emerged as a pivotal paradigm 
for next-generation wireless networks, where dedicated target monitoring terminals (TMTs) can be extensively leveraged for their low-cost flexible deployment and capability to facilitate bistatic and multistatic sensing.
Nevertheless, the coordinated beamforming design for networked ISAC tailored for time-of-arrival (ToA)-based multi-TMT localization remains largely unexplored.
To address this gap, we present a comprehensive study in this paper.
Specifically, we first establish signal models for both communication and localization, and, for the first time, derive a closed-form Cramér-Rao lower bound (CRLB) to quantify the localization performance.
Leveraging this CRLB, we formulate two optimization problems focusing on sensing-centric and communication-centric criteria, respectively, to thoroughly investigate the fundamental communication-localization trade-offs.
For the sensing-centric problem, we develop a globally optimal algorithm based on semidefinite relaxation (SDR), applicable to scenarios where the number of BS antennas exceeds the total number of communication users. In parallel, for the communication-centric problem, we design a globally optimal algorithm for the single-BS case utilizing bisection search. To address the general cases of both problems, we propose a unified and efficient successive convex approximation (SCA)-based algorithm, which is further extended to multi-target scenarios.
Finally,   simulation results demonstrate the effectiveness of our proposed algorithms, reveal the intrinsic trade-offs between communication and localization, and further show that deploying more TMTs is more beneficial than deploying more BSs in networked ISAC systems.
\end{abstract}



\section{Introduction}
\IEEEPARstart{T}{he} proliferation of emerging applications, such as Internet-of-Everything (IoE), 
autonomous driving, and smart cities, 
has created an urgent demand for
wireless networks that are capable of supporting
extensive services
with diverse requirements, including  high data rates,
ultra-reliable low-latency communication,  
and high-accuracy sensing 
\cite{YangPrivacySecurityTrustworthiness2025, xuReconfiguringWireless2023}.
To satisfy these stringent requirements, 
next-generation wireless   networks
are expected to  offer enhanced  flexibility, efficiency, and intelligence
compared to  existing solutions \cite{xuNewPathIntegrated2026, XuDisentangledRepresentationLearning2024}.
Indeed, one promising paradigm for achieving these goals  
is   integrated sensing and communication (ISAC) 
\cite{liuIntegratedSensingCommunications2022}.
In contrast to traditional designs that isolate sensing and communication functionalities,
ISAC   systems    effectively
share  limited  resources such as antennas, spectrum, and power, 
thereby  significantly    reducing deployment costs and improving system efficiency  
\cite{SturmWaveformDesignSignal2011}.

To fully realize the potential of ISAC, 
substantial research efforts have been dedicated to single-base station (BS) scenarios.
A key   focus has been on  transmission   
design targeting both communication-related and sensing-related metrics, 
such as the  signal-to-interference-plus-noise ratio (SINR) 
\cite{huaOptimalTransmitBeamforming2023, heBeamformingOptimizationMultiuser2026},  
energy efficiency (EE) \cite{ZouEnergyEfficientBeamforming2024}, 
beampattern matching error \cite{liuJointTransmitBeamforming2020}, 
and the Cramér-Rao lower bound (CRLB) \cite{liuCramerRaoBoundOptimization2022}. 
Furthermore, the authors in \cite{heUnlockPotential2024} demonstrated that 
integrating ISAC with near-field communication
enables ultra-precise sensing, while the synergy between semantic communication, wireless sensing, and edge learning was explored in \cite{xuEdgeLearningB5G2023}.
Concurrently, researchers have investigated   
pulse waveform designs \cite{xiaoWaveformDesignPerformance2022} and 
 beamforming algorithms \cite{heFullDuplexCommunicationISAC2023a}
to mitigate self-interference  in full-duplex ISAC systems.
Additionally, hybrid  beamforming designs were proposed 
in \cite{singhMultiBeamObjectLocalization2025} to enhance
sensing pattern gains for object localization while guaranteeing communication SINR.
However, single-BS ISAC systems suffer from inherent 
limitations in  coverage, capacity, 
and sensing accuracy,
which may hinder their ability to meet the stringent  requirements of 
next-generation wireless applications \cite{xuTowardsUbiquitous2023}.

Motivated by the success of multi-BS cooperative communication \cite{gesbertMultiCellMIMOCooperative2010} and distributed multiple-input multiple-output (MIMO) radar \cite{godrichTargetLocalizationAccuracy2010}, networked ISAC systems have attracted considerable attention as a promising solution for future wireless networks \cite{zhangTargetLocalizationCooperative2024, chengOptimalCoordinatedTransmit2024}. 
Regarding the transmission design for networked integrated communication and target detection, the authors in \cite{behdadMultiStaticTargetDetection2024} proposed a power allocation scheme to optimize the sensing SINR while ensuring communication quality of service (QoS).
Furthermore, beamforming designs were developed in \cite{chenFastFractionalProgramming2024,babuPrecodingMultiCellISAC2024} to optimize the weighted performance metrics of both communication and parameter estimation.
Turning to integrated communication and target localization, coordinated power control strategies were investigated in \cite{wangConstrainedUtilityMaximization2021, huangCoordinatedPowerControl2022,cuiEnergyEfficientIntegratedSensing2025}
to optimize diverse performance objectives while guaranteeing localization accuracy.
In addition,  beamforming designs for this scenario have  been extensively explored in \cite{gaoCooperativeISACDirect2023, xuSensingassistedRobustSWIPT2025, yangCoordinatedTransmitBeamforming2024}.
For instance, the authors in \cite{gaoCooperativeISACDirect2023} characterized the Pareto boundary of the networked ISAC performance region via coordinated beamforming.

However, these studies predominantly employ BSs to perform sensing reception tasks, configuring them either as dual-functional transceivers or dedicated sensing receivers. The dependence on such stationary infrastructure restricts deployment flexibility, while dedicating full-scale BSs exclusively to sensing results in significant resource underutilization. In contrast, the authors in \cite{xiePerceptiveMobileNetwork2022,xieCollaborativeSensingPerceptive2023} advocate for the use of target monitoring terminals (TMTs) as dedicated sensing receivers. The cost-effectiveness and flexibility of TMTs facilitate dense deployment near targets, enabling multi-view monitoring \cite{xiePerceptiveMobileNetwork2022}. This approach effectively  overcomes the spatial rigidity of fixed BSs and enhances sensing performance through increased spatial diversity. 
Nevertheless, TMTs are designed as low-complexity devices \cite{xieCollaborativeSensingPerceptive2023} and are therefore constrained by a limited number of antennas. This limitation renders the  angle-of-arrival (AoA)-based localization 
adopted in \cite{gaoCooperativeISACDirect2023, 
xuSensingassistedRobustSWIPT2025, 
yangCoordinatedTransmitBeamforming2024} impractical  for multi-TMT localization scenarios,
as  TMTs generally lack   necessary antennas to support 
high-precision AoA  estimation.
To address this, time-of-arrival (ToA)-based localization 
offers a promising solution,
particularly given that clock synchronization between BSs and TMTs is feasibly achievable
\cite{huangCoordinatedPowerControl2022, zhangTargetLocalizationCooperative2024}.



However, the coordinated beamforming design for networked ISAC systems incorporating ToA-based multi-TMT localization
remains unexplored in published literature,
underscoring the need for further investigation.
Firstly, the CRLB   for ToA-based multi-TMT localization is fundamentally different from 
that for AoA-based localization or estimation adopted
in \cite{chenFastFractionalProgramming2024,babuPrecodingMultiCellISAC2024, 
wangConstrainedUtilityMaximization2021,
huangCoordinatedPowerControl2022,cuiEnergyEfficientIntegratedSensing2025,
gaoCooperativeISACDirect2023,xuSensingassistedRobustSWIPT2025, yangCoordinatedTransmitBeamforming2024}, 
necessitating a new theoretical derivation.
Furthermore, given the newly derived
CRLB, a critical open question is how to 
design coordinated beamforming to optimize localization accuracy while 
ensuring  communication QoS,
and vice versa, how to optimize  communication performance while maintaining
satisfactory localization accuracy.
Lastly, distinct from AoA-based models, the CRLB for ToA-based localization 
typically exhibits a  diagonal structure \cite{godrichTargetLocalizationAccuracy2010, lehmannHighResolutionCapabilities2006}. Exploiting this structural property to develop suboptimal or even globally optimal beamforming algorithms remains an open  challenge.
These challenges motivate the present study, in which we make the following contributions.
\begin{itemize}
\item We establish communication and sensing signal 
models for the considered  networked ISAC system.
In particular, 
we derive a closed-form CRLB  for quantifying the performance of target localization,
which serves as a basis for formulating two  different  beamforming design problems:
(i) a sensing-centric problem and (ii) a communication-centric problem.
\item   {For the sensing-centric problem, 
we develop a semidefinite relaxation (SDR)-based
algorithm for the special case where the number of antennas at 
each BS exceeds the total number of communication users (CUs).
Notably, we theoretically analyze the tightness of   SDR,
demonstrating that the proposed 
SDR-based algorithm can achieve the globally optimal solution
with probability one. }
\item { For the communication-centric problem, we propose a 
bisection search-based method for the single-BS case, which is theoretically 
proven to yield the globally optimal solution with probability one.
}
\item {For the general case of both problems, we propose a unified
successive convex approximation (SCA)-based algorithm,
which yields suboptimal performance with low   complexity,
and further extend it from single-target scenarios to more practical multi-target scenarios.}
\item We conduct  extensive  
simulations to validate the
effectiveness of our  proposed algorithms.
Furthermore, we analyze  the performance
trade-offs between communication 
and localization, and demonstrate that deploying more TMTs is more 
beneficial than deploying additional BSs in networked ISAC systems.
\end{itemize}

{
The main differences between this paper and existing works on networked ISAC 
are summarized in Table~\ref{tab:comparison}  for clarity.}

The remainder of this paper is organized as follows. Section~\ref{section2} establishes the signal models and performance metrics for both communication and localization, and formulates the coordinated beamforming design problems. Section~\ref{section3} develops the beamforming algorithms tailored for these problems. Subsequently, Section~\ref{sec:multi_target} extends the proposed SCA-based algorithm to practical multi-target scenarios. Section~\ref{sec:simulation} presents numerical simulation results to validate the proposed algorithms. Finally, Section~\ref{section6} concludes the paper and discusses future research directions.

\subsubsection*{Notations}
$\mathbb{R}$ and $\mathbb{C}$ denote the sets of real and complex numbers, respectively, while $\jmath$ represents the imaginary unit. 
The operators $(\cdot)^T$, $(\cdot)^H$, $(\cdot)^*$, $\mathrm{tr}(\cdot)$, and $\mathrm{diag}(\cdot)$ represent the transpose, Hermitian transpose, complex conjugate, trace, and diagonal, respectively. 
$\mathbf{I}_{M}$ is the $M \times M$ identity matrix. The notation $ \succeq $ ($\succ$) indicates positive semidefinite (positive definite) for matrices and component-wise inequality for vectors.
The Kronecker product is denoted by $\otimes$, while $\mathbf{\Pi}_{\mathbf{A}} = \mathbf{A} ( \mathbf{A}^H \mathbf{A} )^{-1} \mathbf{A}^H$ is the orthogonal projection onto the column space of $\mathbf{A}$. 
Furthermore, $\|\cdot\|$ denotes the Euclidean norm of a vector, and $|\cdot|$ is the modulus of a scalar.
$\mathrm{Re}(\cdot)$, $\mathrm{span}(\cdot)$, and $\partial(\cdot)$ denote the real part, linear span, and partial derivative, respectively. 
Finally, $\log_2(\cdot)$ and $\ln(\cdot)$ denote the base-2 and natural logarithms, $\mathcal{O}(\cdot)$ is the big-O notation, and standard set notations $\in$, $\notin$, $\bigcup$, and $\{ \mathbf{A}_{m,k} \}$ are used.

\begin{table*}[!t]	
\caption{Comparison of This Work with Existing Works on Networked ISAC Systems.}
\label{tab:comparison}
\centering
\renewcommand{\arraystretch}{1.1}
\begin{tabular}{|c|c|c|c|c|c|c|c|c|c|c|c|}
\hline
\textbf{} 
& \cite{behdadMultiStaticTargetDetection2024}
& \cite{chenFastFractionalProgramming2024}
& \cite{babuPrecodingMultiCellISAC2024}
& \cite{wangConstrainedUtilityMaximization2021}  
& \cite{huangCoordinatedPowerControl2022}  
& \cite{cuiEnergyEfficientIntegratedSensing2025} 
& \cite{gaoCooperativeISACDirect2023}
& \cite{xuSensingassistedRobustSWIPT2025}
& \cite{yangCoordinatedTransmitBeamforming2024}
& \cite{xiePerceptiveMobileNetwork2022}
& \textbf{This paper} \\ \hline
\textbf{Coordinated beamforming}  &  & \checkmark & \checkmark & &  & & \checkmark 
	& \checkmark & \checkmark & \checkmark & \checkmark \\ \hline
\textbf{Communication centric}   &  &  & & \checkmark &  & & \checkmark 
	& & \checkmark & & \checkmark \\ \hline
\textbf{Sensing centric}  & \checkmark &  &  & &  & & \checkmark 
	& \checkmark & & & \checkmark \\ \hline
\textbf{SINR as QoS}   & \checkmark  &  & \checkmark & & \checkmark & & 
	& \checkmark & & \checkmark & \checkmark \\ \hline
\textbf{CRLB as QoS}   &  & \checkmark & \checkmark & \checkmark & \checkmark & \checkmark & \checkmark 
	& \checkmark & \checkmark & & \checkmark \\ \hline
\textbf{ToA-based localization}   &  &  & & \checkmark & \checkmark & \checkmark & & 
	& \checkmark & & \checkmark \\ \hline
\textbf{TMTs as sensing receivers}   &  &  & & &  & & & & & \checkmark & \checkmark \\ \hline 
\textbf{SDR-based globally optimal solutions}   &  &  & & &  & & & & & & \checkmark \\ \hline
\textbf{Bisection search-based globally optimal solutions}   &  &  & & &  & & & & & & \checkmark \\ \hline
\end{tabular}  
\end{table*}

\section{System Model and Problem Formulation}
\label{section2}
\begin{figure}[t]
	\centering
    \includegraphics[width=0.49\textwidth]{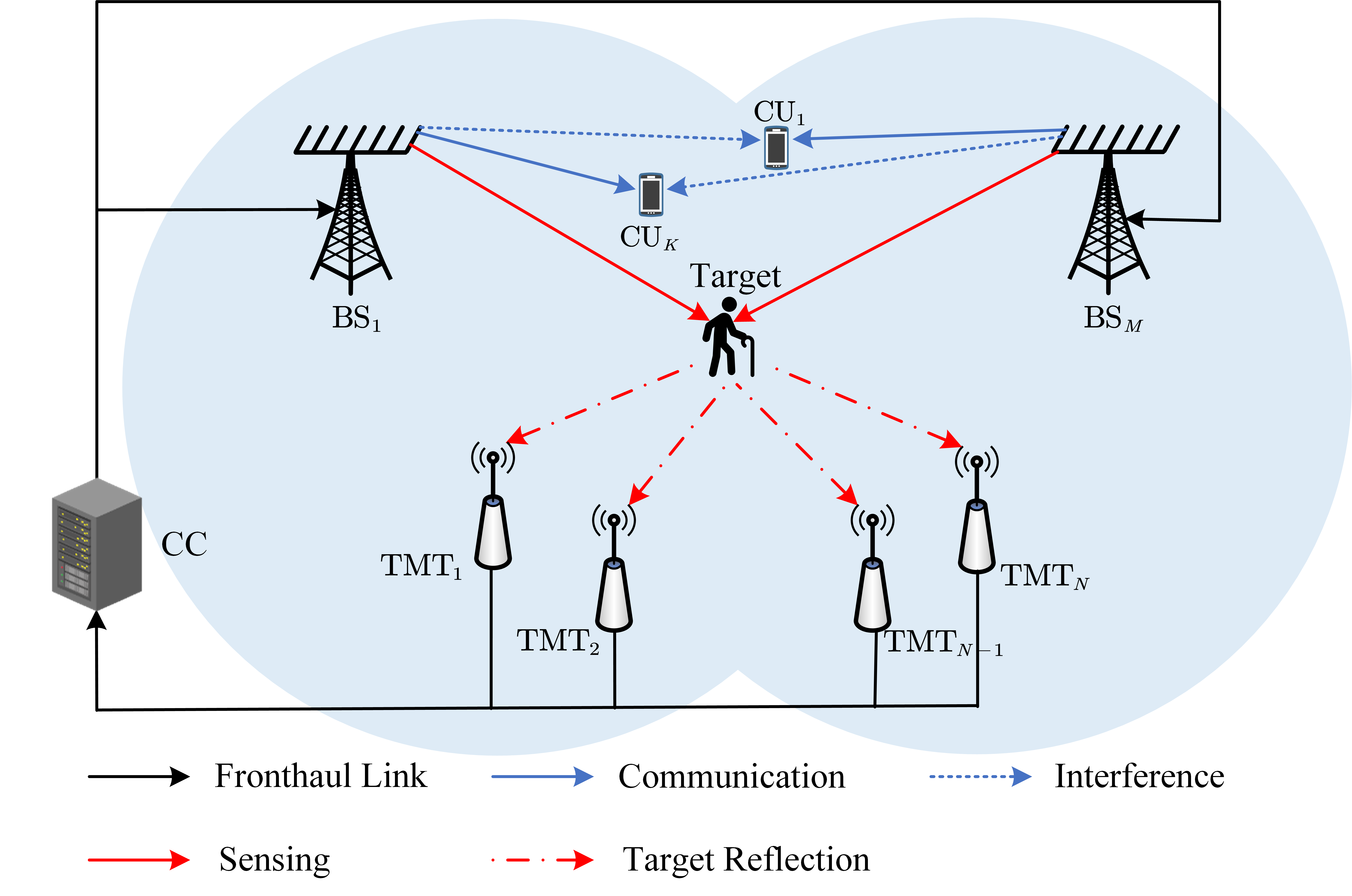}
	\caption{A networked ISAC system with multiple BSs and TMTs.}
	\label{figsetup}
\end{figure}

As shown in Fig. \ref{figsetup}, we consider a networked ISAC system  comprising $M$ BSs, 
$N$ TMTs, 
a central controller (CC), $K$ single-antenna CUs per BS, and a sensing target.
Each BS is equipped with $N_{\mathrm{t}}$ transmitting antennas, while 
each TMT has a single receiving antenna\footnote{ {
    The single-antenna TMT setting does not limit generality, as multiple antennas can be employed to enhance localization performance at the cost of AoA estimation and receiver beamforming, with the proposed algorithms remaining applicable.
}}.
In this architecture, both the BSs and TMTs are connected to the CC via fronthaul links to facilitate data exchange and establish clock synchronization \cite{huangCoordinatedPowerControl2022, zhangTargetLocalizationCooperative2024}.
Specifically, TMTs are envisioned as portable, nomadic sensing nodes (e.g., sensors or radar units) that utilize accessible wired infrastructure interfaces to ensure reliable interconnection with other network elements.
In terms of operation, the BSs transmit ISAC signals to simultaneously serve their associated CUs and illuminate the target.
The TMTs, which are deployed near\footnote{
    Leveraging coarse a priori location knowledge, TMTs operate in a quasi-static manner to cover specific regions of interest, rather than performing continuous real-time target tracking. Redeployment occurs solely when new monitoring tasks emerge in different areas.  This region-centric strategy positions TMTs in significantly closer proximity to potential targets than distant BSs, thereby enhancing localization accuracy while maintaining deployment flexibility and mitigating the prohibitive overhead associated with frequent physical relocation.
}
the target in a distributed manner,
are tasked with collecting the reflected signals and forwarding them to the CC for target localization.
Finally, the CC coordinates the ISAC transmission and processes the sensing data to estimate the target's location. 
This architecture can be deployed over existing cellular networks to monitor the trajectories of vulnerable low-speed targets, such as the elderly or children, thereby enhancing public safety search and rescue capabilities \cite{3gppTR22837},
{albeit with practical overheads associated with TMT placement, synchronization, and coordination.}

\subsection{Transmit Model}
The signal sent by the $m$-th BS, $m \in \left\{1,\cdots,M\right\}$, to its $k$-th CU,
$k \in \left\{1,\cdots,K\right\}$, can be expressed as 
\begin{equation}
    s_{m,k}(t)=\sum_{l=1}^{L} b_{m,k,l}g(t-lT_{\mathrm{s}}).
\end{equation}
Here, $L$, 
$T_{\mathrm{s}}$, and $b_{m,k,l} \in \mathbb{C}$ denote the total number of sensing  snapshots, 
the duration of each symbol,  and the   transmitted information symbol
of the $m$-th BS for serving its $k$-th CU 
at the $l$-th snapshot, $l \in \left\{1,\cdots,L\right\}$, respectively. 
The duration of the 
ISAC period of interest is given by $T = L T_{\mathrm{s}}$.
Besides, $g(t)$ denotes the   real-valued 
baseband pulse signal which  
satisfies the following two properties  \cite{niuTargetLocalizationTracking2012}
\begin{subequations}
    \begin{align}
        \frac{1}{T_{\mathrm{s}}}\int_{0}^{T} \left|g(t)\right|^{2}\mathrm{d}t&=1, \\
        \int_{-\infty}^{\infty} f\left|G(f)\right|^{2}\mathrm{d}f&=0,
    \end{align}
\end{subequations} 
where $G(f)$ is the 
Fourier transform of $g(t)$. 

The transmitted information symbols $\left\{b_{m,k,l}\right\}$ are assumed to be 
independent and identically distributed (i.i.d.) 
random variables with zero-mean and   unit variance.
Consequently, when $L$ is sufficiently large, 
the transmitted signals become asymptotically 
mutually orthogonal with unit average power \cite{chengOptimalCoordinatedTransmit2024,
huangCoordinatedPowerControl2022}, 
i.e.,
\begin{subequations}
    \begin{align}
        \int_0^{T} s_{m , i}^*(t) s_{n, j}(t-\tau) \mathrm{d} 
        t&=0, \ \forall \tau, (m,i) \neq (n,j),  \\
        \frac{1}{T}\int_0^{T} \left| s_{m , i}(t) \right|^2  \mathrm{d} 
        t&=1, \ \forall m, i.
    \end{align} \label{orthogonal}
\end{subequations}

\subsection{Communication Model}
Let $\mathbf{h}_{i, m, k} \in \mathbb{C}^{N_{\mathrm{t}} \times 1}$ 
denote the  communication channel from the
$i$-th BS, $i \in \left\{1,\cdots,M\right\}$, to the $k$-th user   
served by  the $m$-th BS.
The channel can be characterized by the multi-path channel model
as  
\begin{equation}
    \begin{aligned}
        \mathbf{h}_{i, m, k}= & \sqrt{\frac{1}{V}}
    \sum_{v = 1}^{V} \alpha_{i, m, k, v} \mathbf{a}
    \left(\phi_{i, m, k, v}\right),
    \end{aligned} %
\end{equation}
where $V$ represents the total number of  paths. 
The parameters $\alpha_{i, m, k, v} \in \mathbb{C}$ and $\phi_{i, m, k, v}$ represent  the channel   coefficient 
and the angle of departure (AoD) for the $v$-th  path  
from the $i$-th BS to the $k$-th CU  served by the $m$-th BS, 
respectively. 
Furthermore, the vector $\mathbf{a}\left(\phi\right) \in \mathbb{C}^{N_{\mathrm{t}} \times 1}$ 
denotes the transmitting array response vector with respect to the AoD $\phi$.
Within the considered scenario, each BS is equipped with a uniform linear array (ULA). 
The transmitting array response vector in terms of AoD $\phi$
can be expressed as 
\begin{equation}
    \mathbf{a}\left(\phi\right)=\left[1, e^{\jmath 2 \pi \frac{d}{\lambda} \sin \left(\phi\right)}, \ldots,
    e^{\jmath 2 \pi \frac{d}{\lambda}\left(N_{\mathrm{t}}-1\right) \sin \left(\phi\right)}\right]^T,
\end{equation}
where $d$ and $\lambda$ denote the inter-antenna spacing and
the wavelength of the carrier frequency, respectively.

At time instance $t$, the  signal received by  the $k$-th CU  associated with the $m$-th BS is 
expressed as
\begin{equation}
    \begin{aligned}
        \hat{s}_{m,k}(t)  = & \ \mathbf{h}_{m, m, k}^H \mathbf{f}_{m, k} s_{m,k}(t) \\ & + \sum\limits_{(i,j) \neq (m,k)}   \mathbf{h}_{i, m, k}^H \mathbf{f}_{i, j} s_{i, j}(t)  + z_{m,k}(t), \label{5}
    \end{aligned} 
\end{equation}
where $z_{m,k}(t)$ is the circularly symmetric complex Gaussian (CSCG) noise
with zero-mean and variance $\sigma_{\mathrm{n}}^2$, 
and $\mathbf{f}_{m,k} \in \mathbb{C}^{N_{\mathrm{t}} \times 1}$  denotes  
the $k$-th beamforming vector associated 
with  the  $m$-th BS.
Without loss of generality,  the SINR
is chosen as  the performance metric for  communication. From   \eqref{5},  
the SINR of the $k$-th CU  served by the  $m$-th BS  is given by
\begin{equation}
    \label{eq-sinr}
    \mathrm{SINR}_{m, k}=\frac{\left|\mathbf{h}_{m, m, k}^H 
    \mathbf{f}_{m, k}\right|^2}{\sum\limits_{(i,j) \neq (m,k)} \left|\mathbf{h}_{i, m, k}^H \mathbf{f}_{i, j}\right|^2+
    \sigma_{\mathrm{n}}^2}.
\end{equation}

\subsection{Sensing Model}
The signals transmitted by the   BSs are reflected by the target
and subsequently captured by the TMTs. 
At a given time instance $t$, the  signal received by the  $n$-th TMT,
$n \in \left\{1,\cdots,N\right\}$,
can be mathematically expressed as\footnote{
In addition to the desired reflection path from the BSs to the target and subsequently to the TMTs, interfering signals may significantly degrade localization performance. 
In practice, interference at the TMTs can be classified into two categories: (i) the direct path from the BSs and (ii) scattered paths from surrounding clutter, including CUs. 
Since the direct path is generally stable over time, it can be reliably estimated and cancelled prior to localization~\cite{behdadMultiStaticTargetDetection2024}. 
Regarding clutter-induced interference, various identification and mitigation techniques are available,  
including machine learning (ML)-based approaches~\cite{huangMachineLearningEnabledLOS2020}, which leverage the robust classification capabilities of ML models to detect and suppress such signals. 
Consequently, consistent with existing networked ISAC studies~\cite{chenFastFractionalProgramming2024, babuPrecodingMultiCellISAC2024, wangConstrainedUtilityMaximization2021, huangCoordinatedPowerControl2022}, we assume that interference is effectively mitigated and focus exclusively on the desired sensing path.
}
\begin{equation}
    \begin{aligned}
        \hspace{-2mm} r_n(t) \! &= \! \underbrace{
        \sum_{m=1}^M \! \sum_{k=1}^K \varepsilon_{m, n}  
        \mathbf{a}^H \! \left(\theta_m\right) \! \mathbf{f}_{ m, k}  s_{m, k}\! \left(t\! - \!\tau_{m, n}\right)}_{\mu_n(t)}   +   n_n(t), 
    \end{aligned}
\end{equation}
where $\tau_{m, n} \in \mathbb{R}$ represents the propagation delay of the ISAC 
signal from the  $m$-th BS to  the $n$-th TMT,  
$\varepsilon_{m, n} \in \mathbb{C}$ refers to the channel coefficient  incorporating 
both the large-scale channel fading coefficient and the  radar cross section (RCS), and  
$\theta_m$ represents the AoD 
from  the $m$-th BS to the target.
Additionally, $\mu_n(t)$ and $n_n(t)$ represent the useful signal for localization 
and the CSCG noise, respectively.

At the CC, the aggregation of signals from the $N$ TMTs
can be mathematically represented as 
\begin{equation}
    \begin{aligned}
        \mathbf{r}(t)&=\left[
    r_1(t), \ldots, r_{N}(t)\right]^T \\ &=\left[
        {\mu}_1(t), \ldots, {\mu}_{N}(t)\right]^T+\mathbf{n}_{\mathrm{s}}(t).
    \end{aligned}
    \label{r}
\end{equation}
Here, $\mathbf{n}_{\mathrm{s}}(t)$ denotes the CSCG   
noise, which is both spatially and temporally white,  
with zero-mean and  autocorrelation function $\sigma_{\mathrm{s}}^2 \mathbf{I}_N
\delta(\tau)$, 
where 
$\sigma_{\mathrm{s}}^2$ and $\delta(\tau)$, respectively, denote the
power spectral density (PSD) of   noise and the Dirac delta function.
Owing to the orthogonality 
among   transmitted  signals as shown 
in \eqref{orthogonal},
the CC can accurately estimate time delays from variations
in the envelope of   transmitted signals. 
Subsequently, it  determines the  location of the target  by 
employing the well-known ToA-based localization method  \cite{godrichTargetLocalizationAccuracy2010}.
Specifically, once the time delays have been estimated, 
the location of the  target can be determined by   a set 
of equations,  
which are expressed as 
{
\begin{equation}
    \label{tau}
    \begin{aligned}
    \tau_{m,n} = \ & \frac{1}{\mathrm{c}}\Big(\sqrt{\left(x_{m}-x\right)^2+\left(y_{m}-y\right)^2 + H^2 }  \\
    & +\sqrt{\left(x^{\prime}_{n}-x\right)^2+\left(y^{\prime}_{n}-y\right)^2}\Big), \ \forall m, n.
    \end{aligned}%
\end{equation}
Here, $\mathrm{c}$ denotes the speed of light, and $H$ represents the height of BSs. The TMTs and the target are assumed to be located on the ground plane with zero height \cite{guJointlyOptimizeThroughput2024}. %
The parameters $x_{m}$ and $y_{m}$ denote the $x$- and $y$-coordinates of the  $m$-th BS, 
$x^{\prime}_{n}$ and $y^{\prime}_{n}$ represent  the  coordinates 
of the $n$-th TMT, and $x$ and $y$ correspond to the coordinates of  the target.}


To evaluate  the   sensing performance, 
the  CRLB for  target localization
is adopted as the performance metric. The detailed derivation of the CRLB is presented below.

\begin{theorem}
    \label{theorem1}
    The sum of  CRLB for the ToA-based estimation of the target's location $\left(x,y\right)$ 
    from \eqref{r} and \eqref{tau}
    is given by
    \begin{equation}
        \label{CRLBxy}
        {C}_{x,y}  =
        \mathrm{tr} \Big( \Big(\boldsymbol{\Lambda} \mathbf{Z} \boldsymbol{\Lambda}^T\Big)^{-1} \Big),
    \end{equation}
    where 
    \begin{equation}
        \label{Lambda}
        \setlength{\arraycolsep}{1pt}
        \begin{aligned}
        &\boldsymbol{\Lambda} =
        \left[\begin{array}{cccc}
        \frac{\partial}{\partial x} \boldsymbol{\uptau}^T \\
        \frac{\partial}{\partial y} \boldsymbol{\uptau}^T \\
        \end{array}\right]
        \end{aligned}
    \end{equation}
    denotes the Jacobian matrix which is
    composed of   partial derivatives $\frac{\partial}{\partial x} \boldsymbol{\uptau}^T$ 
    and $\frac{\partial}{\partial y} \boldsymbol{\uptau}^T$, and
    \begin{equation}
        \label{Z}
        \begin{aligned}
            \mathbf{Z} = \mathrm{diag}& \left(J(\tau_{1,1},\tau_{1,1}),\ldots,J(\tau_{M,N},\tau_{M,N})\right)
        \end{aligned}
    \end{equation}
    signifies the Fisher information submatrix  related to  the time delays, and 
    $J(\tau_{m,n},\tau_{m,n})$ denotes the element   with respect to the time delay $\tau_{m,n}$, 
    expressible as
    \begin{equation}
        \begin{aligned}
            J(\tau_{m,n},\tau_{m,n})  = \ & \frac{8\pi^2 T  \beta^2  \left| \varepsilon_{m,n} \right|^2}
            {\sigma_{\mathrm{s}}^2}  \\ 
            & \times  \mathbf{a}^H\left(\theta_m\right) 
            \Big( 
            \sum_{k=1}^{K}\mathbf{f}_{m,k} \mathbf{f}_{m,k}^{H} \Big)  \mathbf{a}\left(\theta_m\right),
            \label{J_ele}
        \end{aligned} 
    \end{equation}
    where $\beta=\sqrt{\frac{\int_{-\infty}^{\infty} f^2 \left|G(f)\right|^2 \mathrm{d}f}
    {\int_{-\infty}^{\infty} \left|G(f)\right|^2 \mathrm{d}f} }$ is the effective bandwidth 
    of the pulse signal $g(t)$ \cite{godrichTargetLocalizationAccuracy2010,huangCoordinatedPowerControl2022}.
\end{theorem}
\begin{IEEEproof}
    The proof is provided in Appendix \ref{app:theorem1}.
\end{IEEEproof}


\subsection{Problem Formulation} 
Based on the parameters $\left\{\theta_m\right\}$ and $\left\{\varepsilon_{m,n}\right\}$ estimated in the preceding ISAC period, we perform coordinated beamforming optimization 
for the subsequent transmission phase.
Specifically, we formulate two distinct optimization problems: a sensing-centric formulation and a communication-centric formulation. 
Notably, these two formulations effectively characterize the complete trade-off profile between communication and localization \cite{gaoCooperativeISACDirect2023}. Furthermore, the proposed SCA-based algorithm is generic and can be extended to address the weighted-sum joint design problem with minor modifications. Therefore, we focus on these two representative cases to maintain clarity and avoid redundancy.

First, we consider the sensing-centric design, which prioritizes the localization performance by minimizing the CRLB while satisfying a communication QoS requirement. 
The corresponding optimization problem is formulated as
\begin{subequations}
    \begin{align}
        \mathop{\text{minimize}}\limits_{\left\{\mathbf{f}_{m, k}\right\}} \quad & C_{x, y} \\
        \text {subject to} \quad 
        &\sum_{k=1}^{K} \left\|\mathbf{f}_{m, k}\right\|^{2} \leq P, \ \forall m,  \label{1b} \\
        &\mathrm{SINR}_{m, k}  \geq \eta, \ \forall m, k. \label{1c}
    \end{align} \label{eq:problem}%
\end{subequations}
Constraint \eqref{1b} imposes the maximum transmit power budget $P$ for each BS. 
Furthermore, \eqref{1c} ensures the communication QoS requirement for each CU, where $\eta$ denotes the minimum SINR threshold.

Second, we address the communication-centric design. This formulation aims to maximize the minimum SINR among all CUs subject to a predefined localization accuracy requirement. 
The problem is mathematically expressed as
\begin{subequations}
    \begin{align}
        \mathop{\text{maximize}}\limits_{\left\{\mathbf{f}_{m, k}\right\}} \quad & \min_{m,k} 
        \left\{\mathrm{SINR}_{m, k} \right\} \\
        \text {subject to} \quad
        &\sum_{k=1}^{K} \left\|\mathbf{f}_{m, k}\right\|^{2} \leq P, \ \forall m, \label{2b}   \\
        &C_{x, y} \leq \epsilon, \label{2c}
    \end{align} \label{eq:problem2}%
\end{subequations}
where \eqref{2c} enforces the sensing QoS requirement, with $\epsilon$ representing the maximum tolerable CRLB threshold.



\section{Proposed Solutions for Networked ISAC}
\label{section3}
In this section, we develop efficient algorithms for problem~\eqref{eq:problem} and problem~\eqref{eq:problem2}, respectively.

\subsection{Sensing-centric Problem}
The optimization problem in \eqref{eq:problem} is nonconvex,  particularly due 
to the nonconvex nature of $C_{x,y}$ with respect to $\left\{\mathbf{f}_{m,k}\right\}$, 
stemming from its fractional structure and the presence of quadratic terms. To address this challenge, we first reformulate problem \eqref{eq:problem} into a 
more tractable form.
Specifically, we   introduce a set of auxiliary optimization variables, 
namely $\left\{q_m\right\}$,
and then reformulate problem \eqref{eq:problem}  by leveraging the 
diagonal structure of $\mathbf{Z}$,
as discussed  below.
\begin{lemma}
	\label{lemma2}
	The problem in \eqref{eq:problem} is equivalent to  the following problem
	\begin{subequations}
		\begin{align}
			\mathop{\text{minimize}}\limits_{{\left\{\mathbf{f}_{m, k}, \ q_m\right\}}} 
			\quad & \mathrm{tr} 
			\Big( \Big(\boldsymbol{\Lambda} \hat{\mathbf{Z}} 
			\boldsymbol{\Lambda}^T\Big)^{-1} \Big)  \label{ob} \\
			\text {subject to} \quad &  \eqref{1b}, \ \eqref{1c},         \\
			& \begin{aligned}
				q_m  \leq \ & \mathbf{a}^H\left(\theta_m\right) 
				\Big( 
				\sum_{k=1}^{K}\mathbf{f}_{m,k}\mathbf{f}_{m,k}^H \Big) \\
				& \times \mathbf{a}\left(\theta_m\right), \ \forall m,
			\end{aligned} \label{qqq} \\
			& q_m \geq 0, \ \forall m. \label{q_nonge}
		\end{align} \label{eq:problem_q}%
	\end{subequations}
	Here, $\hat{\mathbf{Z}}$ is a diagonal 
	matrix, which is defined as
	\begin{subequations}
		\begin{align}
			\hat{\mathbf{Z}} & =  \mathrm{diag} 
				\Big(q_1 \hat{\mathbf{Z}}_1,   \ldots,
					q_M \hat{\mathbf{Z}}_M \Big), \label{hatZ} \\
			\hat{\mathbf{Z}}_m & =   \frac{8\pi^2 T 
			\beta^2}{\sigma_{\mathrm{s}}^2} \mathrm{diag} 
				\Big( \left| \varepsilon_{m,1} \right|^2,   \ldots,
					\left| \varepsilon_{m,N} \right|^2 \Big). \label{hatZm}
		\end{align}
	\end{subequations}
\end{lemma}
\begin{IEEEproof}
	The proof is provided in Appendix \ref{app:lemma1}.
\end{IEEEproof}

{Although problem \eqref{eq:problem_q} remains nonconvex due to 
SINR constraint   \eqref{1c} and   newly introduced constraint   \eqref{qqq}, 
the quadratic terms have been removed from the fractional objective, offering greater 
flexibility for subsequent
algorithm design. 
Specifically, when the condition $N_{\mathrm{t}} > MK$ is satisfied, 
the problem reduces to a special case for which we develop a globally 
optimal solution based on the SDR technique. While for the general case, we 
propose a suboptimal yet efficient algorithm based on the SCA technique.}
{ \begin{remark}
	For a multi-antenna BS, the number of antennas $N_{\mathrm{t}}$ is 
	typically larger than the number of served CUs $K$ in order to 
	satisfy communication QoS requirements, especially in millimeter-wave 
	(mmWave) systems \cite{bjornsonMassiveMIMOSub6GHz2019}.
	Therefore, in small- to medium-scale networks 
	with a moderate number of cooperative BSs $M$, the 
	product $MK$ remains relatively small, making it 
	highly likely that $N_{\mathrm{t}} > MK$ holds.
\end{remark}}

\subsubsection{Special case of $N_{\mathrm{t}} > MK$}
In this case, we employ the SDR technique to solve problem \eqref{eq:problem_q} optimally.
Specifically, by defining $\mathbf{F}_{m,k} = \mathbf{f}_{m,k}\mathbf{f}_{m,k}^H, \ \forall m,k$, 
and temporarily relaxing the rank-one constraints, 
{problem \eqref{eq:problem_q} can be reformulated by 
replacing $\left\{\mathbf{f}_{m,k}\mathbf{f}_{m,k}^H\right\}$ with $\left\{\mathbf{F}_{m,k}\right\}$.}
The resulting problem is expressed as 
\begin{subequations}
    \begin{align}
        \mathop{\text{minimize}}\limits_{{\left\{\mathbf{F}_{m, k}, \ q_m\right\}}} \ & \mathrm{tr} 
			\Big( \Big(\boldsymbol{\Lambda} \hat{\mathbf{Z}} \boldsymbol{\Lambda}^T\Big)^{-1} \Big) \\
		\text {subject to} \ &\mathrm{tr}\Big(\sum_{k=1}^K \mathbf{F}_{m, k} \Big)
		\leq P, \ \forall m, \label{SDPb}  \\
        &              \begin{aligned} 
							& \eta \Big( \sum_{(i,j) \neq (m,k)}  \mathbf{h}_{i, m, k}^H 
							\mathbf{F}_{i, j} \mathbf{h}_{i, m, k} +
							\sigma_{\mathrm{n}}^2 \Big) \\
							& \leq \mathbf{h}_{m, m, k}^{H} \mathbf{F}_{m, k} \mathbf{h}_{m, m, k}, \ \forall m, k,
						\end{aligned} \label{SDPc} \\
		& q_m   \leq \mathbf{a}^H\left(\theta_m\right) \Big( 
		\sum_{k=1}^{K}\mathbf{F}_{m,k} \Big)   \mathbf{a}\left(\theta_m\right), \  \forall m, \label{SDPd} \\ 
		& q_m \geq 0, \ \forall m, \label{q_nonge2}  \\
		&  \mathbf{F}_{m,k} \succeq \mathbf{0}, \ \forall m, k, \label{SDPe}  
    \end{align}
    \label{eq:problem_eq}%
\end{subequations}
which is a convex problem and   can therefore be solved in polynomial time 
with CVX \cite{grantCVXMatlabSoftware2014}.

The relaxation of  rank-one constraints on $\left\{\mathbf{F}_{m,k}\right\}$
in   problem \eqref{eq:problem_eq} may yield solutions that are not rank-one,
potentially resulting in suboptimal solutions for   original problem   \eqref{eq:problem}. 
{However, when $N_{\mathrm{t}} > MK$, the optimal solutions of 
problem \eqref{eq:problem_eq} are guaranteed to be rank-one with probability one.}
This implies that the optimal solutions of original problem \eqref{eq:problem} can be directly obtained by applying 
eigenvalue decomposition to the solutions of problem \eqref{eq:problem_eq}.
This   is presented in the following theorem and analysis.
{
\begin{theorem}
	\label{theorem2}
	Provided  that problem \eqref{eq:problem_eq} is feasible.
	If $\mathbf{a}\left(\theta_m\right) \notin \mathrm{span}
	\Big( \bigcup_{i,j} \mathbf{h}_{m,i,j} \Big), \ \forall m$, then
	the optimal solutions 
	of   problem \eqref{eq:problem_eq}
	are  rank-one with probability one.
\end{theorem}
\begin{IEEEproof}
	The proof is provided in Appendix \ref{app:theorem2}.
\end{IEEEproof}
}

Note that
$\mathrm{span}\left( \bigcup_{i,j} \mathbf{h}_{m,i,j} \right)$
has a dimension of at most $MK$ when $N_{\mathrm{t}} > MK$,
while $\mathbf{a}(\theta_m)$ lies in an $N_{\mathrm{t}}$-dimensional space.
Therefore, the probability that $\mathbf{a}(\theta_m)$ lies entirely in $\mathrm{span}\left( \bigcup_{i,j} \mathbf{h}_{m,i,j} \right)$ approaches zero when $N_{\mathrm{t}} > MK$.
Therefore, when $N_{\mathrm{t}} > MK$, 
the optimal solutions of problem \eqref{eq:problem_eq} are guaranteed to be rank-one with probability one, implying that the optimal solutions of original problem \eqref{eq:problem} can be obtained with probability one.

\subsubsection*{Complexity Analysis}  
The complexity of this SDR-based algorithm is about
$\mathcal{O}\left( \ln(1/\varsigma)  M^{3.5} K^{3.5}N_{\mathrm{t}}^{6.5}  \right)$ 
\cite{heEnergyEfficientBeamforming2022}, where 
$\varsigma > 0$ is a predefined solution accuracy.
{\begin{remark}
	Although this SDR-based algorithm incurs high computational complexity, 
	it can deliver the globally optimal solution,
	which is a rare and valuable property for nonconvex optimization problems.
	When the number of   antennas $N_{\mathrm{t}}$
	is small or moderate, we can directly apply 
	this algorithm to solve problem \eqref{eq:problem}. Otherwise, this algorithm 
	may not be practical due to its high computational complexity, but it can still 
	serve as a benchmark for evaluating the performance of 
	other more efficient algorithms thanks to its optimality.
\end{remark}}

{
\subsubsection{General case} \label{sec:communication_sensing_sca}
The SDR-based algorithm guarantees optimality  with probability one
only when $N_{\mathrm{t}} > MK$, and its high computational complexity 
further limits its practical applicability. 
To address this, we propose a suboptimal yet 
computationally efficient algorithm based on 
the SCA technique for the general case.}

{
First, we reformulate problem \eqref{eq:problem_q} into the following form
\begin{subequations}
    \begin{align}
		\hspace{-0.1cm}
		\underset{ \substack{\left\{\mathbf{f}_{m, k}, \ \rho_{m,k}\right\}
		\\ \left\{ u_{m,k}, \ q_m\right\}} }{\text{minimize}}  \quad &
					\mathrm{tr} 
		\Big( \Big(\boldsymbol{\Lambda} \hat{\mathbf{Z}} 
		\boldsymbol{\Lambda}^T\Big)^{-1} \Big) \\
        \text {subject to }   \quad & \eqref{1b}, \  \eqref{qqq}, \ \eqref{q_nonge}, \\
        &\frac{\rho_{m,k}^2}{u_{m,k}} \geq \eta, \ \forall m,k, \label{21d}  \\
		&\mathbf{h}_{m, m, k}^H 
			\mathbf{f}_{m, k} \mathbf{f}_{m, k}^H \mathbf{h}_{m,   m,   k}     \geq     
			\rho_{m,k}^2,  \  \forall m, k,  \label{21e} \\
		&\sigma_{\mathrm{n}}^2 + \omega  \left(  \left\{ \mathbf{f}_{i,j} \right\}, m, k \right)  
		\leq u_{m,k}, \ \forall m, k,   \label{21f}
    \end{align} \label{eq:problem_sensing_sca}%
\end{subequations}
where $\left\{\rho_{m,k}\right\}$ and $\left\{u_{m,k}\right\}$ are the newly introduced auxiliary variables,
and 
\begin{equation}
	\omega  \left(  \left\{ \mathbf{f}_{i,j} \right\}, m, k \right)   =    \sum_{(i,j) \neq (m,k)}  
	\mathbf{h}_{i, m, k}^H \mathbf{f}_{i, j} \mathbf{f}_{i, j}^H  \mathbf{h}_{i, m, k}, \ \forall m, k.
\end{equation}
It is clear that the equivalence between problem \eqref{eq:problem_q} and problem \eqref{eq:problem_sensing_sca} is ensured. Now, 
the nonconvexity of   problem   \eqref{eq:problem_sensing_sca} 
is mainly due to the constraints in \eqref{qqq}, \eqref{21d}, and \eqref{21e}.
Fortunately,  we can handle these constraints by applying the SCA technique.}

First, by defining $\mathbf{f}_{m}   = \left[ \mathbf{f}_{m,1}^T, \ldots, 
\mathbf{f}_{m,K}^T \right]^T, \ \forall m$,
the constraint in \eqref{qqq} can be rewritten as 
\begin{equation}
	\begin{aligned}
		\mathbf{f}_{m}^H   \mathbf{D}_{m}  \mathbf{f}_{m} \geq q_m, \ \forall m,
	\end{aligned}  \label{sub1_11}	
\end{equation}
where we define the positive 
semidefinite matrix $\mathbf{D}_{m}$ as 
\begin{equation}
	\mathbf{D}_{m} = \mathbf{I}_K \otimes 
		\left(  \mathbf{a}\left( \theta_m \right) \mathbf{a}^H\left( \theta_m \right) \right), \ \forall m.
\end{equation}
The constraint in \eqref{sub1_11} is still nonconvex,
but the SCA technique can be exploited to establish its convex subset 
by using the first-order Taylor series expansion.
Specifically, given   solutions $\left\{\mathbf{f}_{m}^{(r)}\right\}$ 
obtained at the  $r$-th iteration of SCA,
the  surrogate  constraint for \eqref{sub1_11} 
at the $\left( r+1 \right)$-th iteration can 
be established as \cite{yangCoordinatedTransmitBeamforming2024}
\begin{equation}
	\begin{aligned}
		2\operatorname{Re}\Big( \mathbf{f}_{m}^H   \mathbf{D}_m  \mathbf{f}_{m}^{(r)} \Big)
		- \left(\mathbf{f}_{m}^{(r)}\right)^H   \mathbf{D}_m  \mathbf{f}_{m}^{(r)}  \geq q_m, \ \forall m,
	\end{aligned}  \label{sub1_1_21}
\end{equation}
which is an affine constraint and can be efficiently processed.
Similarly, given    solutions  
$\left\{\mathbf{f}_{m,k}^{\left(r\right)}\right\}$, 
$\left\{\rho_{m,k}^{\left(r\right)}\right\}$, and $\left\{u_{m,k}^{\left(r\right)}\right\}$
obtained at the $r$-th iteration of SCA,
the surrogate constraint for \eqref{21d} at the $\left( r+1 \right)$-th 
iteration can be established as \cite{yangCoordinatedTransmitBeamforming2024}
\begin{equation}
	\begin{aligned}
		\frac{2\rho_{m,k}^{\left(r\right)}}{u_{m,k}^{\left(r\right)}} \rho_{m,k}-
		\frac{\left(\rho_{m,k}^{\left(r\right)}\right)^2}{\left(u_{m,k}^{\left(r\right)}\right)^2} u_{m,k}
		\geq \eta, \ \forall m, k,
	\end{aligned} \label{sub1_1_1_21}
\end{equation}
and the surrogate constraint for
\eqref{21e} at the $\left( r+1 \right)$-th iteration can be established as \cite{yangCoordinatedTransmitBeamforming2024}
\begin{equation}
	\begin{aligned}
		& 2\operatorname{Re} \left( \mathbf{f}_{m,k}^H   \mathbf{h}_{m,m,k} \mathbf{h}_{m,m,k}^H  
		\mathbf{f}_{m,k}^{(r)} \right) \\ 
		& - \left(\mathbf{f}_{m,k}^{(r)}\right)^H   \mathbf{h}_{m,m,k} \mathbf{h}_{m,m,k}^H  
		\mathbf{f}_{m,k}^{(r)} 
		\geq  \rho_{m,k}^2, \ \forall m,k.
	\end{aligned}  \label{sub1_1_31}
\end{equation}

{
Building on the above surrogate constraints, 
the problem in \eqref{eq:problem_sensing_sca}
at the $\left( r+1 \right)$-th iteration of SCA can be formulated as 
\begin{subequations}
    \begin{align}
		\hspace{-0.1cm}
        \mathop{\text{maximize}}\limits_{\substack{\left\{\mathbf{f}_{m, k}, \ \rho_{m,k}\right\}
		\\ \left\{ u_{m,k}, \ q_m\right\}}} \quad
					& \mathrm{tr} 
		\Big( \Big(\boldsymbol{\Lambda} \hat{\mathbf{Z}} 
		\boldsymbol{\Lambda}^T\Big)^{-1} \Big) \\
        \text {subject to } \quad
        &  \eqref{1b}, \ \eqref{sub1_1_21}, \ \eqref{q_nonge}, \ \eqref{sub1_1_1_21}, \ \eqref{sub1_1_31}, \ \eqref{21f}, 
    \end{align} \label{eq:problem23333}%
\end{subequations}
which is  a convex problem and thus
can be   efficiently solved by CVX \cite{grantCVXMatlabSoftware2014}.
The  problem in  \eqref{eq:problem_sensing_sca} can be addressed
by iteratively solving a sequence of    problem    \eqref{eq:problem23333} 
until convergence.}

{The initial point can be 
obtained by first finding a feasible solution 
to problem \eqref{eq:problem}, denoted as $\mathbf{f}_{m,k}^{(0)}$, 
and then substituting it into \eqref{21e} and \eqref{21f} to obtain $\rho_{m,k}^{(0)}$ 
and $u_{m,k}^{(0)}$, respectively.
The corresponding feasibility problem can be optimally solved using 
existing methods \cite{wieselLinearPrecodingViaConic2006}, thereby guaranteeing that a 
feasible initial point is always available whenever problem \eqref{eq:problem} is feasible.
This is crucial for the convergence and stability of the SCA-based algorithm, 
as an infeasible initial point can lead to infeasible surrogate problems and impede convergence.
}

The  overall algorithm to address  problem   \eqref{eq:problem} 
for the general case  is
presented in Algorithm~\ref{alg:SCA}, with its convergence property summarized in
the following corollary.
{\begin{corollary}
	\label{corollary1}
	In sensing-centric scenarios, the sequence of objective values generated by Algorithm \ref{alg:SCA} is guaranteed to converge. Provided that Slater's condition holds for problem \eqref{eq:problem23333} at each iteration, the limit of any convergent subsequence satisfies the Karush-Kuhn-Tucker (KKT) conditions.
\end{corollary}
\begin{IEEEproof}
	The proof is provided in Appendix \ref{app:corollary1}. 
\end{IEEEproof}}

\subsubsection*{Complexity Analysis} 
The complexity of this SCA-based  algorithm is about
$\mathcal{O}\left(   I  \ln(1/\varsigma)  
M^{3.5}K^{3.5}N_{\mathrm{t}}^3    \right)$   \cite{wangOutageConstrainedRobust2014a},
where $I$ is the number of iterations for SCA to converge, and
$\varsigma > 0$ is a predefined solution accuracy\footnote{
	In massive-scale systems where complexity is a bottleneck, gradient-based methods \cite{maiEnergyEfficiencyMaximization2022} offer a promising avenue to further alleviate the computational burden, potentially delivering satisfactory performance with very low overhead.
}.

\subsection{Communication-centric Problem}
Problem   \eqref{eq:problem2} is challenging to solve due to two main reasons.
First, its objective function is nonconvex due to the max-min-ratio    structure.
Second,  constraint  \eqref{2c} is nonconvex, as it contains    quartic terms in a fractional form.
To overcome these challenges, we address problem \eqref{eq:problem2} by 
developing a globally optimal solution for the special case of $M=1$, 
and propose a suboptimal yet efficient solution for the general case.

\subsubsection{Special case of $M=1$}
In this case, we propose a  novel bisection search-based method to optimally solve problem \eqref{eq:problem2}.
To simplify the presentation, we omit the subscript $m$ in the following analysis.

We first reformulate  problem    \eqref{eq:problem2} to  the following form
\begin{subequations}
    \begin{align}
		\hspace{-0.18cm}
        \mathcal{S}\left(P,\epsilon\right): \mathop{\text{maximize}}\limits_{\left\{\mathbf{f}_{k}\right\}, \ q} \quad
					& \min_{k} \left\{ \mathrm{SINR}_{k} \right\} \\
        \text {subject to} \quad
        &  \sum_{k=1}^{K} \left\|\mathbf{f}_{k}\right\|^{2} \leq P,  \label{1c_2} \\
		& \begin{aligned}
			q  \leq \ & \mathbf{a}^H\left(\theta\right) 
			\Big( 
			\sum_{k=1}^{K}\mathbf{f}_{k}\mathbf{f}_{k}^H \Big)  
			\mathbf{a}\left(\theta\right), 
		\end{aligned} \label{qqq_2} \\
		& q \geq 0, \label{q_nonge3} \\
        & \mathrm{tr} 
		\Big( \Big(\boldsymbol{\Lambda} \hat{\mathbf{Z}} \boldsymbol{\Lambda}^T\Big)^{-1} \Big) 
		\leq \epsilon, \label{2c_22}
    \end{align} \label{eq:problem2211}%
\end{subequations}
where $q$ is a newly introduced optimization variable  and
$\mathcal{S}\left(P,\epsilon\right)$ is a function of $P$ and $\epsilon$ 
that maps to the optimal value of the problem in \eqref{eq:problem2211}.
For the case of $M=1$, 
the optimal value and   optimal  solutions of   problem   \eqref{eq:problem2211} are exactly 
identical to those of the original problem in \eqref{eq:problem2}.

Then, we   construct the following power minimization problem  
\begin{subequations}
	\begin{align}
		\hspace{-0.18cm}
		\mathcal{P}\left(\eta, P, \epsilon\right): \mathop{\text{minimize}}
		\limits_{\left\{\mathbf{f}_{k}\right\}, \ q} \quad & \sum_{k=1}^{K} \left\|\mathbf{f}_{k}\right\|^{2}   / P \\
		\text {subject to} \quad  
		& \eqref{qqq_2}, \ \eqref{q_nonge3}, \ \eqref{2c_22},    \\
		& \frac{\left|\mathbf{h}_{ k}^H 
    		\mathbf{f}_{k}\right|^2}{\sum_{j \neq k} \left|\mathbf{h}_{k}^H \mathbf{f}_{j}\right|^2+
    		\sigma_{\mathrm{n}}^2} \geq \eta, \ \forall k,
	\end{align} \label{eq:problem23}%
\end{subequations}
where  $\mathcal{P}\left(\eta, P, \epsilon\right)$ is a function of $\eta$, 
$P$, and $\epsilon$ that maps to the optimal value of the problem in \eqref{eq:problem23}.
Subsequently, we can bridge  $\mathcal{S}\left(P,\epsilon \right)$ and $\mathcal{P}\left(\eta,P,\epsilon\right)$
by the following Lemma.
\begin{lemma}
	\label{lemma3}
	$\mathcal{S}\left(P,\epsilon\right)$ and  $\mathcal{P}\left(\eta,P,\epsilon \right)$ have the following relationships
	\begin{subequations}
		\begin{align}
			\mathcal{S}\left(\mathcal{P}\left( \eta,P,\epsilon \right) P,\epsilon\right) & = \hat{\eta}, \\
			\mathcal{P}\left(\mathcal{S}\left(P,\epsilon\right),P,\epsilon\right) & = 1,
		\end{align}
	\end{subequations}
	where $\hat{\eta}$ is the maximum value in the set 
	$\left\{x | \mathcal{P}\left(x,P,\epsilon\right) = \mathcal{P}\left(\eta,P,\epsilon\right)\right\}$.
\end{lemma}
\begin{IEEEproof}
	The proof is provided in Appendix \ref{app:lemma2}.
\end{IEEEproof}


Lemma \ref{lemma3} reveals that solving problem \eqref{eq:problem2211}
can be transformed into the process of solving problem \eqref{eq:problem23}.
Specifically, due to the non-decreasing nature of $\mathcal{P}\left(\eta,P,\epsilon\right)$ with respect to $\eta$,
we can identify the maximum value of $\eta$  that satisfies  
the condition $\mathcal{P}\left(\eta,P,\epsilon\right) = 1$
by employing the bisection search method.
The obtained $\eta$ and the corresponding solutions $\left\{\mathbf{f}_{k} \right\}$
are the optimal value and the optimal solutions of problem \eqref{eq:problem2211}, respectively.
{While this transformation has long been a powerful tool in 
wireless communications \cite{wieselLinearPrecodingViaConic2006}, we extend 
its applicability to the context of networked integrated communication and 
localization and further exploit it to design a globally optimal algorithm,
which has not been reported in existing networked ISAC studies.
}

Subsequently, we solve problem \eqref{eq:problem23} optimally by leveraging the SDR technique.
Specifically, by defining $\mathbf{F}_{k} = \mathbf{f}_k \mathbf{f}_k^H, \ \forall k$, and temporarily
omitting the rank-one constraints,
problem \eqref{eq:problem23} can be reformulated as
\begin{subequations}
	\begin{align}
		\mathop{\text{minimize}}
		\limits_{\left\{\mathbf{F}_{k}\right\}, \ q} \quad & \mathrm{tr}\Big(\sum_{k=1}^K \mathbf{F}_{k} \Big) / P \\
		\text {subject to} \quad  
		& \mathrm{tr} 
		\Big( \Big(\boldsymbol{\Lambda} \hat{\mathbf{Z}} 
		\boldsymbol{\Lambda}^T\Big)^{-1} \Big) \leq \epsilon, \label{2c_232} \\
		& \begin{aligned} 
			& \eta \Big( \sum_{j \neq k}  \mathbf{h}_{k}^H 
			\mathbf{F}_{j} \mathbf{h}_{k} +
			\sigma_{\mathrm{n}}^2 \Big) \\
			& \leq \mathbf{h}_{k}^{H} \mathbf{F}_{k} \mathbf{h}_{k}, \ \forall k,
		\end{aligned} \label{SDPc_232}  \\
		& q   \leq \mathbf{a}^H\left(\theta\right) \Big( 
		\sum_{k=1}^{K}\mathbf{F}_{k} \Big)   \mathbf{a}\left(\theta\right),   \label{SDPd232} \\ 
		& q \geq 0,  \label{SDPe2321} \\
		&  \mathbf{F}_{k} \succeq \mathbf{0}, \ \forall k, \label{SDPe232}
	\end{align} \label{eq:problem232}%
\end{subequations}
which is a convex problem and thus can be solved in polynomial time
by CVX \cite{grantCVXMatlabSoftware2014}.
The feasibility of problem \eqref{eq:problem232} is guaranteed by the
following lemma.
\begin{lemma}
	\label{lemma4}
	Define $\mathbf{H}=\left[\mathbf{h}_1,\cdots,\mathbf{h}_K\right]$.
	If   $\mathbf{H}$ has full column rank,   problem   \eqref{eq:problem232} is 
	always feasible.
\end{lemma}
\begin{IEEEproof}
	The proof is provided in Appendix \ref{app:lemma3}.
\end{IEEEproof}

Due to the random nature of the channel, 
the matrix $\mathbf{H}$ is almost always full column rank in practice.
Therefore, the feasibility of problem \eqref{eq:problem232} is guaranteed with probability one.
In addition, the tightness of SDR can also be guaranteed in this case, 
which is stated in the following theorem.
{
\begin{theorem}
	\label{theorem3}
	Provided  that   problem   \eqref{eq:problem232} is feasible, 
	the optimal solutions of problem \eqref{eq:problem232}
	are  rank-one with probability one.
\end{theorem}}
\begin{IEEEproof}
	The proof is incorporated into that of Theorem \ref{theorem2}
	and is therefore omitted here for brevity.
\end{IEEEproof}

Theorem \ref{theorem3} indicates that the optimal value of problem \eqref{eq:problem23} is identical to that of problem \eqref{eq:problem232}, and its optimal solutions can be directly obtained via eigenvalue decomposition of the solutions to problem \eqref{eq:problem232}.
The overall algorithm to solve   communication-centric problem    \eqref{eq:problem2} 
for the special case of $M=1$ is
presented in Algorithm \ref{alg:BiSearch},
where $\eta_{\mathrm{up}}$ is a sufficiently 
large value   serving as an 
upper bound   in the bisection search
and $\eta_{\mathrm{tol}}$ is a predefined tolerance. 
We now provide the following convergence guarantee for the entire algorithm.
{
\begin{corollary}
	\label{corollary2}
	Algorithm \ref{alg:BiSearch} is 
	guaranteed to converge to the globally optimal solution of 
	problem \eqref{eq:problem2} with probability one.
\end{corollary}}
\begin{IEEEproof}
	This can be directly derived from Lemma \ref{lemma3}, Lemma \ref{lemma4}, and  Theorem \ref{theorem3}.
\end{IEEEproof}

\subsubsection*{Complexity Analysis} 
The complexity of this   bisection search-based
algorithm is about $\mathcal{O}\left(  \log_2 (\eta_{\mathrm{up}}/\eta_{\mathrm{tol}})  \ln(1/\varsigma)  
K^{3.5}N_{\mathrm{t}}^{6.5}    \right)$ 
\cite{heEnergyEfficientBeamforming2022}, where $\varsigma > 0$ is a predefined solution accuracy.


\begin{algorithm}[t]
	\caption{A Unified Algorithm for Sensing-Centric and 
	Communication-Centric Problems in the General Case}
	\label{alg:SCA}
	\begin{algorithmic}[1]
		\STATE \textbf{Initialization:} Set $r=0$, and 
		initialize $\Big\{\mathbf{f}_{m,k}^{(0)}\Big\}$,
			$\left\{\rho_{m,k}^{(0)}\right\}$, and $\left\{u_{m,k}^{(0)}\right\}$ with feasible solutions corresponding to either sensing-centric or communication-centric scenarios.
		\REPEAT
		\STATE  Given $\left\{\mathbf{f}_{m,k}^{(r)}\right\}$, 
				$\left\{\rho_{m,k}^{(r)}\right\}$, 
				and $\left\{u_{m,k}^{(r)}\right\}$, solve  problem   \eqref{eq:problem23333} 
				for sensing-centric scenarios, or   
				problem  \eqref{eq:problem2333} for communication-centric scenarios,
				to obtain solutions  $\left\{\mathbf{f}_{m,k}^{\star}\right\}$, 
				$\left\{\rho_{m,k}^{\star}\right\}$, and $\left\{u_{m,k}^{\star}\right\}$.
		\STATE  Update $\left\{\mathbf{f}_{m,k}^{(r+1)}\right\} = \left\{\mathbf{f}_{m,k}^{\star} \right\}$,
				$\left\{\rho_{m,k}^{(r+1)}\right\} = \left\{\rho_{m,k}^{\star} \right\}$, and 
				$\left\{u_{m,k}^{(r+1)}\right\} = \left\{u_{m,k}^{\star} \right\}$.
		\STATE  Set $r=r+1$.
		\UNTIL{convergence.}
		\STATE \textbf{Output:} $\left\{\mathbf{f}_{m,k}^{(r)}\right\}$.
	\end{algorithmic}
\end{algorithm}

\begin{algorithm}[t]
	\caption{An Algorithm for Communication-Centric Problem for the Special Case of $M=1$}
	\label{alg:BiSearch}
	\begin{algorithmic}[1]
		\STATE \textbf{Initialization:} Set $\eta_{\mathrm{low}}=0$, $\eta_{\mathrm{up}}$, and $\eta_{\mathrm{tol}}$.
		\REPEAT
		\STATE   Set $\eta = \left( \eta_{\mathrm{low}} + \eta_{\mathrm{up}} \right) / 2$.
		\STATE Solve   problem   \eqref{eq:problem232} with $\eta$ to obtain optimal value $a$,
			   and then obtain optimal solutions $\left\{\mathbf{f}_{k}\right\}$ 
			   by eigenvalue decomposition. 
		\IF{$a \leq 1$}
		\STATE $\eta_{\mathrm{low}} = \eta$.
		\ELSE
		\STATE $\eta_{\mathrm{up}} = \eta$.
		\ENDIF
		\UNTIL{$\eta_{\mathrm{up}} - \eta_{\mathrm{low}} < \eta_{\mathrm{tol}}$.}
		\STATE \textbf{Output:} $\left\{\mathbf{f}_{k}\right\}$.
	\end{algorithmic}
\end{algorithm}

\subsubsection{General case}
Note that   Algorithm \ref{alg:BiSearch} 
can also be applied to the general case   with minor modifications. 
However, the tightness of SDR cannot be guaranteed in general, which implies that convergence of Algorithm~\ref{alg:BiSearch} to the global optimum is not assured. Fortunately, Algorithm~\ref{alg:SCA} remains applicable to problem~\eqref{eq:problem2}.

{
Specifically, problem~\eqref{eq:problem2} can be addressed using the SCA technique, following a similar approach to that in Section~\ref{sec:communication_sensing_sca}.  At the $(r+1)$-th iteration of SCA, 
the surrogate problem for problem~\eqref{eq:problem2} can be formulated as
\begin{subequations}
    \begin{align}
		\hspace{-0.1cm}
        \mathop{\text{maximize}}\limits_{\substack{\left\{\mathbf{f}_{m, k}, \ \rho_{m,k}\right\}
		\\ \left\{ u_{m,k}, \ q_m\right\}, \ \varpi}} \quad & \varpi \\
        \text {subject to} \quad \ \
        &  \eqref{2b}, \ \eqref{sub1_1_21}, \ \eqref{q_nonge}, \ \eqref{sub1_1_31}, \ \eqref{21f}, \\
		& \mathrm{tr} 
		\Big( \Big(\boldsymbol{\Lambda} \hat{\mathbf{Z}} \boldsymbol{\Lambda}^T\Big)^{-1} \Big) 
		\leq \epsilon, \label{2c_2} \\
		& \begin{aligned}
		& \frac{2\rho_{m,k}^{\left(r\right)}}{u_{m,k}^{\left(r\right)}} \rho_{m,k}-
			\frac{\left(\rho_{m,k}^{\left(r\right)}\right)^2}
			{\left(u_{m,k}^{\left(r\right)}\right)^2} u_{m,k} \\
				& \geq \varpi, \ \forall m, k,
		\end{aligned} \label{32c}
    \end{align} \label{eq:problem2333}%
\end{subequations}
where $\varpi$ is a newly introduced optimization variable.
It can be observed that problem~\eqref{eq:problem2333} exhibits a structure similar to that of problem~\eqref{eq:problem23333}.
Accordingly, Algorithm~\ref{alg:SCA} serves as a unified framework that can be directly applied to both sensing-centric and communication-centric problems in the general case.
}

{
The complexity analysis for communication-centric scenarios is the same as that for sensing-centric scenarios, and the convergence behavior in communication-centric scenarios is summarized in the following corollary.
{\begin{corollary}
	\label{corollary3}
	In communication-centric scenarios, the sequence of objective values generated by Algorithm \ref{alg:SCA} is guaranteed to converge. Provided that Slater's condition holds for problem \eqref{eq:problem2333} at each iteration, the limit of any convergent subsequence satisfies the KKT conditions.
\end{corollary}
\begin{IEEEproof}
	The proof is provided in Appendix \ref{app:corollary3}.
\end{IEEEproof}}

{
\section{Extension to Multi-Target Scenarios}
\label{sec:multi_target}
In this section, we extend the system model from   
single-target   to   multi-target scenarios 
with $U$ targets indexed by $u \in \left\{1, \ldots, U\right\}$.
We consider two types of multi-target cases: i)    widely separated targets, 
and ii)     closely spaced targets.
In the first case, where the targets are widely separated, 
it is feasible to adopt different groups of TMTs 
to sense different targets, with each group dedicated to a 
specific one. Under this setting, the sensing tasks for different targets 
can be treated as independent. Accordingly,  the CRLB for the 
$u$-th target retains the same form as in   single-target scenarios, and is given by
\begin{equation}
	\label{CRLBxyu}
	{C}_{u}  =
	\mathrm{tr} \left( \left(\boldsymbol{\Lambda}_u \mathbf{Z}_u 
	\boldsymbol{\Lambda}_u^T\right)^{-1} \right), \ \forall u,
\end{equation}
where $\boldsymbol{\Lambda}_u$ and $\mathbf{Z}_u$ are defined 
similarly to \eqref{CRLBxy}, but indexed by 
$u$ to indicate the corresponding target.
}

{
In the second   case, where the targets are closely spaced,  
the TMTs are jointly used to sense all targets.
A key challenge in this case is that the signals reflected from 
different targets are mutually coupled, making it 
necessary for the TMTs to associate the received signals with the corresponding targets.
This  is known as the data association problem, 
a fundamental issue in multi-target 
localization \cite{shenMultipleSourceLocalization2014}.
If the data association problem is solved optimally, the CRLB for each 
target remains the same as in \eqref{CRLBxyu} 
\cite{shenMultipleSourceLocalization2014}. 
Even when this problem is not perfectly resolved, \eqref{CRLBxyu}   still serves 
as a valid performance bound for each target.
Therefore, we adopt  \eqref{CRLBxyu} as a unified CRLB expression  
for  multi-target scenarios.}

{
Based on this unified expression, the sensing-centric optimization problem in \eqref{eq:problem} 
can be extended to multi-target scenarios  by adopting a min-max criterion, leading to the following problem
\begin{subequations}
    \begin{align}
        \mathop{\text{minimize}}\limits_{\left\{\mathbf{f}_{m, k}\right\}, \ \omega} \quad & \omega \\
        \text {subject to} \quad & \eqref{1b}, \ \eqref{1c}, \\
		& C_{u} \leq \omega, \ \forall u. \label{CRLB_multi} 
    \end{align} \label{eq:problem_multi}%
\end{subequations}
The communication-centric   problem in \eqref{eq:problem2} becomes
\begin{subequations}
    \begin{align}
        \mathop{\text{maximize}}\limits_{\left\{\mathbf{f}_{m, k}\right\}} \quad & \min_{m,k} 
        \left\{\mathrm{SINR}_{m, k} \right\} \\
        \text {subject to} \quad & \eqref{2b}, \\
        &C_{u} \leq \epsilon,  \ \forall u. \label{CRLB_multi2}
    \end{align} \label{eq:problem_multi2}%
\end{subequations}
It is worth noting that Algorithm~\ref{alg:SCA} can be directly applied to address 
both \eqref{eq:problem_multi} and \eqref{eq:problem_multi2}, 
as their mathematical structures remain consistent 
with those in   single-target scenarios.}

\section{Numerical Simulations}
\label{sec:simulation}
In this section, we present numerical simulations to evaluate the proposed algorithms.

\subsection{Simulation Setup and Parameter Settings}
In the simulations, we consider a networked ISAC system 
comprising $M=2$ BSs,  $N=4$ TMTs, and $K=4$ CUs per BS. 
The BSs are positioned at coordinates $(80,80\sqrt{3})$m and $(80,-80\sqrt{3})$m, respectively. 
The target is located at the origin for simplicity.
The TMTs are placed at $(50,50)$m, $(50,-50)$m, $(-50,50)$m, and $(-50,-50)$m, respectively.
The coordinates of CUs within the area are generated randomly.


The sensing channel coefficient is modeled as $\varepsilon_{m, n} = \sqrt{F_{m,n}}\zeta_{m,n}$, where $F_{m,n} = \frac{\mathrm{c}^2}{f_{\mathrm{c}}^2(4\pi)^3 d_{m}^2 (d^{\prime}_{n})^2}$ denotes the large-scale fading coefficient between the $m$-th BS and the $n$-th TMT. Here, $d_m$ and $d^{\prime}_n$ represent the distances from the target to the $m$-th BS and the $n$-th TMT, respectively, and $f_{\mathrm{c}}$ is the carrier frequency. The term $\zeta_{m,n}$ characterizes the RCS associated with the $m$-th BS and the $n$-th TMT, which is assumed to be a zero-mean  Gaussian random variable with unit variance~\cite{chengOptimalCoordinatedTransmit2024,behdadMultiStaticTargetDetection2024}. Regarding the communication links, the   channel coefficient is expressed as $\alpha_{i, m, k, v } = \sqrt{\tilde{F}_{i,m,k}}\tilde{\zeta}_{i,m,k,v}$. The large-scale fading coefficient is defined as $\tilde{F}_{i,m,k} = \frac{\mathrm{c}^2}{f_{\mathrm{c}}^2(4\pi)^2 \tilde{d}_{i,m,k}^2}$, where $\tilde{d}_{i,m,k}$ denotes the distance between the $i$-th BS and the $k$-th CU served by the $m$-th BS. The small-scale fading term $\tilde{\zeta}_{i,m,k,v}$ follows a standard complex Gaussian distribution~\cite{kwonIntegratedLocalizationCommunication2023}. 

Throughout the simulations, the parameters are set as follows: the number of   paths is $V = 10$, the BS height is $20$~m, and the maximum transmit power is $P = 30$~dBm. Additional simulation details are provided in Table~\ref{tab:simulation_parameters}.

\begin{table}[t]
	\centering
	\caption{Simulation Parameters}
	\begin{tabular}{|c|c|c|}
		\hline
		\textbf{Notation} & \textbf{Value} & \textbf{Description} \\ \hline
		$f_{\mathrm{c}}$ & 24 GHz \cite{SturmWaveformDesignSignal2011} & Carrier frequency \\ \hline
		$N_{\mathrm{t}}$ & {32} & Number of antennas at each BS \\ \hline
		$\beta$ & 100 MHz \cite{SturmWaveformDesignSignal2011} & Effective bandwidth \\ \hline
		$\sigma_{\mathrm{n}}^2$ & -94 dBm & Power of communication noise\\ \hline
		$\sigma_{\mathrm{s}}^2$ & -174 dBm/Hz \cite{huangCoordinatedPowerControl2022} & PSD of sensing noise\\ \hline
		$L$ & 256 \cite{SturmWaveformDesignSignal2011} & Total number of sensing snapshot \\ \hline
	\end{tabular}
	\label{tab:simulation_parameters}
\end{table}

{
\subsection{Benchmark}
To evaluate the performance of the proposed algorithms, we compare them against several benchmark algorithms.
\begin{itemize}
	\item \textbf{Radar-only:} This benchmark removes   communication constraint
	\eqref{1c} from the sensing-centric optimization problem \eqref{eq:problem},
	which serves as a performance upper bound in   sensing-centric scenarios.
	\item \textbf{Communication-only:} This benchmark  removes  sensing constraint
	\eqref{2c}	from the communication-centric optimization problem \eqref{eq:problem2}, 
	which serves as a performance upper bound in   communication-centric scenarios.
	\item \textbf{Zero-forcing (ZF) \cite{behdadMultiStaticTargetDetection2024,
	chengOptimalCoordinatedTransmit2024}:}  
	This benchmark employs ZF beamforming
	as the baseline method to tackle both the sensing-centric and communication-centric problems.
	Define $\mathbf{H}_{m,m,k} = \left[ 
	\mathbf{h}_{m,1,1}, \ldots, \mathbf{h}_{m,m,k-1}, \mathbf{h}_{m,m,k+1}, \ldots, \mathbf{h}_{m,M,K}\right]$.
	Its singular value decomposition (SVD) is given by $\mathbf{H}_{m,m,k} = 
	\left[\mathbf{U}_{m,m,k}, \tilde{\mathbf{U}}_{m,m,k}\right] \boldsymbol{\Sigma}_{m,m,k} \mathbf{V}_{m,m,k}^H$.
	where $\tilde{\mathbf{U}}_{m,m,k} \in \mathbb{C}^{N_t \times (N_t-MK+1)}$ is the 
	orthogonal complement of $\mathbf{U}_{m,m,k} \in \mathbb{C}^{N_t \times (MK-1)}$.
	Then, the ZF beamforming vector   is given by
	$
	\mathbf{f}_{m, k}^{\mathrm{ZF}}=
		\frac{\sqrt{p_{m, k}^{\mathrm{ZF}}} \tilde{\mathbf{U}}_{m, m, k}
			 \tilde{\mathbf{U}}_{m, m, k}^{H} 
				\mathbf{h}_{m, m, k}}{\left\| \tilde{\mathbf{U}}_{m, 
					m, k} \tilde{\mathbf{U}}_{m, 
					m, k}^{H} \mathbf{h}_{m, m, k}\right\|}, \ \forall m, k,
	$
	where $p_{m, k}^{\mathrm{ZF}}$ is the power allocated to the $k$-th CU at the $m$-th BS.
	These vectors are then incorporated into   the two optimization problems, 
	yielding convex power allocation problems that can be solved optimally. 
	\item \textbf{Minimum mean square error (MMSE):} This benchmark employs MMSE beamforming
	as the baseline method to tackle both the sensing-centric and communication-centric problems. 
	Specifically, the  MMSE beamforming vectors can be obtained by 
	solving the following   problem
	\begin{equation}
	\begin{aligned}
		&\underset{\left\{\mathbf{f}_{m,k}, \ \alpha_m\right\}}{\text{minimize}} \quad 
			\sum_{m=1}^{M}\sum_{k=1}^{K}\left\| \mathbf{s}_{m,k} - \sqrt{\alpha_m} \hat{\mathbf{s}}_{m,k} \right\|^2 \\
		&\text{subject to} \quad \sum_{k=1}^{K} \left\| \mathbf{f}_{m,k} \right\|^2 \leq P, \ \forall m,
	\end{aligned}
	\end{equation}
	where $\alpha_m$ is the scaling factor for the $m$-th BS.
	Although this problem is nonconvex, it can be addressed by the
	alternating optimization   \cite{heFullDuplexCommunicationISAC2023a}.
	The obtained solutions 
	are expressed as $\left\{\hat{\mathbf{f}}_{m,k}\right\}$.
	Then, the MMSE beamforming vector   is given by
	$
	\mathbf{f}_{m,k}^{\mathrm{MMSE}} = \frac{\sqrt{p_{m,k}^{\mathrm{MMSE}}}\hat{\mathbf{f}}_{m,k}}
		{ \left\|\hat{\mathbf{f}}_{m,k}\right\|}, \ \forall m, k,
	$
	where $p_{m,k}^{\mathrm{MMSE}}$ is the power allocated to the $k$-th CU at the $m$-th BS.
	These vectors are then incorporated into the two optimization problems, 
	yielding convex power allocation problems that can be solved optimally.
	\item \textbf{Beampattern matching \cite{liuJointTransmitBeamforming2020}:}  
	This benchmark employs beampattern matching algorithm
	to tackle the sensing-centric problem. Specifically, the beamforming vectors can be obtained by
	solving the following   problem
	\begin{equation}
	\begin{aligned}
		&\underset{\left\{\mathbf{R}_m, \ \mathbf{f}_{m,k}, \ \alpha_m\right\}}{\text{minimize}} \quad 
			\sum_{m=1}^{M}\sum_{l=1}^{L} R_{m,l} \\
		& \ \ \ \text{subject to} \quad \quad \sum_{k=1}^{K} \left\| \mathbf{f}_{m,k} \right\|^2 = P, \ \forall m,
	\end{aligned} \label{eq:beampattern}%
	\end{equation}
	where $R_{m,l}$ is defined as 
	$\left| \alpha_m d \left(\vartheta_l^m\right)
	- \mathbf{a}^H \left(\theta_m\right) 
				\left(\sum_{k=1}^{K}\mathbf{f}_{m,k}\mathbf{f}_{m,k}^H + \mathbf{R}_m \right)  
				\mathbf{a}\left(\theta_m\right) \right|^2$
	with $\alpha_m$ is the scaling factor, $d\left(\vartheta_l^m\right)$ is the desired 
	beampattern at the angle $\vartheta_l^m$, and $\mathbf{R}_m$ is the covariance 
	matrix of the dedicated
	sensing signal for the $m$-th BS.
	The angle ${\vartheta_l^m}$ is obtained 
	by uniformly sampling $[-90^\circ,90^\circ]$ 
	with $L = 360$ points. The desired beampattern of the $m$-th BS is given by
	\begin{equation}
	d\left(\vartheta_l^m\right) = \begin{cases}
		1, & \text{if } \theta_m -  5^\circ \leq \vartheta_l^m \leq \theta_m + 5^\circ, \\
		0, & \text{otherwise.}
	\end{cases}
	\end{equation} 
	This problem can be solved optimally by using 
	the method proposed in \cite{liuJointTransmitBeamforming2020}.
\end{itemize}
}

\subsection{Performance Evaluation}
\subsubsection{Sensing-centric Beamforming Algorithms}
We first evaluate   the  
proposed sensing-centric beamforming algorithms
against   benchmark algorithms.

Fig. \ref{figCRLBvsSINR} evaluates the CRLB performance  
as a function of   SINR threshold $\eta$. 
The results reveal that
the proposed   algorithms exhibits superior performance 
compared to the ZF, MMSE, and 
beampattern matching algorithms. 
Moreover, an increase in the SINR threshold   leads 
to a higher CRLB for the proposed algorithms. 
This occurs because a more stringent SINR threshold 
forces the ISAC system to direct the transmitted beam 
more towards CUs, leading to less power
directed at   the target.  This observation 
underscores the intrinsic  
trade-offs between   localization and communication,
where an improvement in communication performance may result in a degradation of localization accuracy.
{Additionally, Algorithm~\ref{alg:SCA} achieves near-optimal performance compared with the SDR-based algorithm, 
demonstrating its capability to effectively balance computational complexity and performance. 
This also suggests that the SDR-based algorithm serves as a suitable benchmark for evaluating suboptimal algorithms.}

\begin{figure}[t]
	\centering
	\includegraphics[width=3.3in]{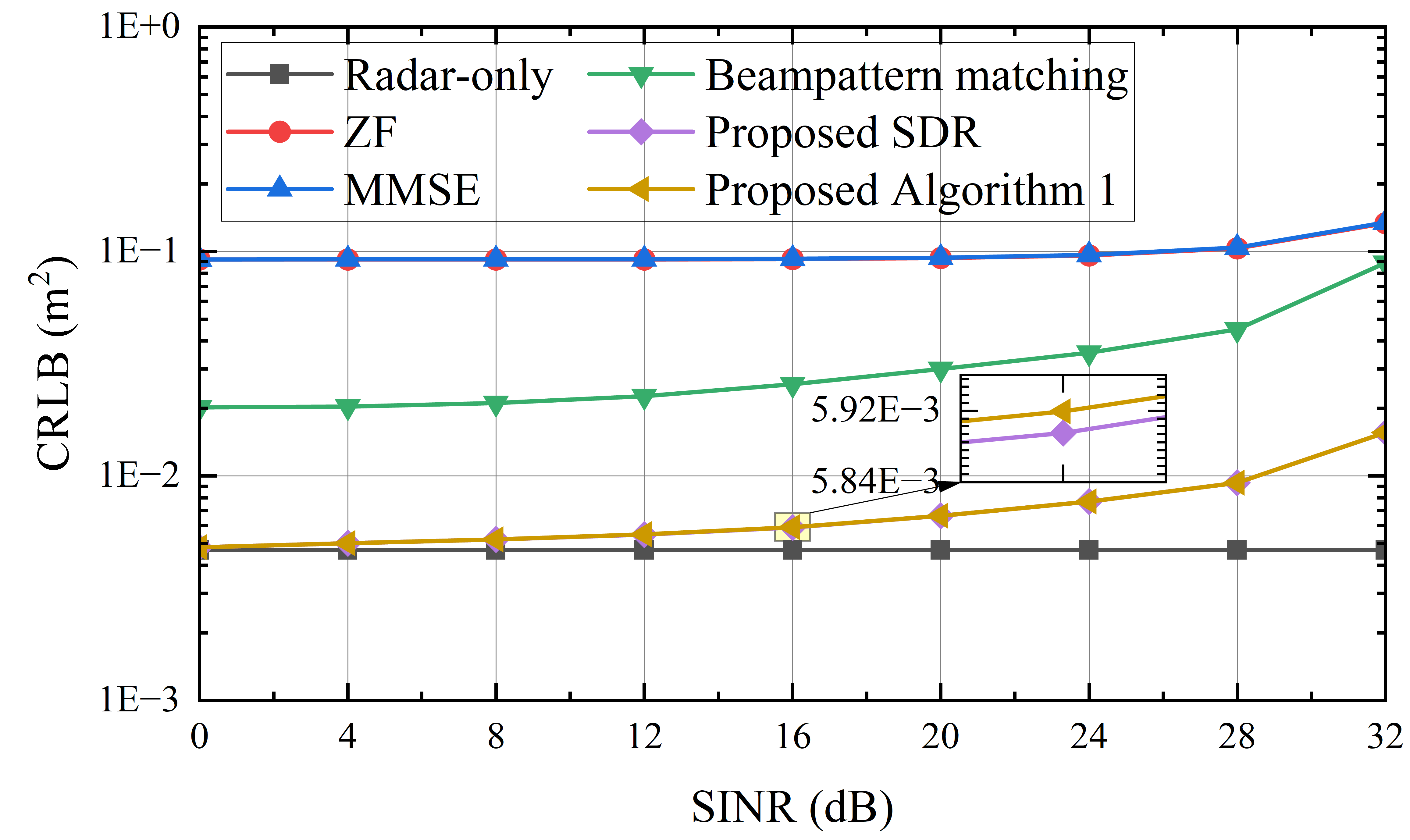}
	\caption{CRLB performance versus $\eta$.}
	\label{figCRLBvsSINR}
\end{figure}

The resultant beampatterns of all the considered algorithms are depicted in Fig. \ref{figbeampattern}.
In this simulation, the target is positioned at angles of $60^\circ$ and $-60^\circ$ relative 
to both BSs, respectively.  
We observe that the proposed  algorithm   generate 
the desired beampatterns, with main lobes precisely  aligned at $60^\circ$ and $-60^\circ$ 
for both BSs, mirroring the performance of  the radar-only    algorithm. 
In contrast, the beampattern matching algorithm also aligns its 
main lobes at these angles  
but yields less sharp main lobes compared with the proposed algorithm. 
{This is because the CRLB minimization criterion can be 
viewed as implicitly maximizing the beampattern gain in the target direction, 
whereas the beampattern matching criterion focuses on minimizing the 
deviation between the designed and desired beampatterns, 
potentially compromising the beampattern gain to achieve 
a closer overall fit. As a result, the former 
yields sharper beampatterns toward the target direction.}
Moreover, both the ZF  and  MMSE algorithms fail to form directional beams 
towards the target directions for both BSs, 
highlighting their inadequacy for radar sensing applications.
These simulation results are consistent  
with the CRLB results in Fig. \ref{figCRLBvsSINR},
further demonstrating 
the superior performance of the proposed   algorithm.

\begin{figure}[t]
\centering
\subfloat[]{\includegraphics[width=3.3in]{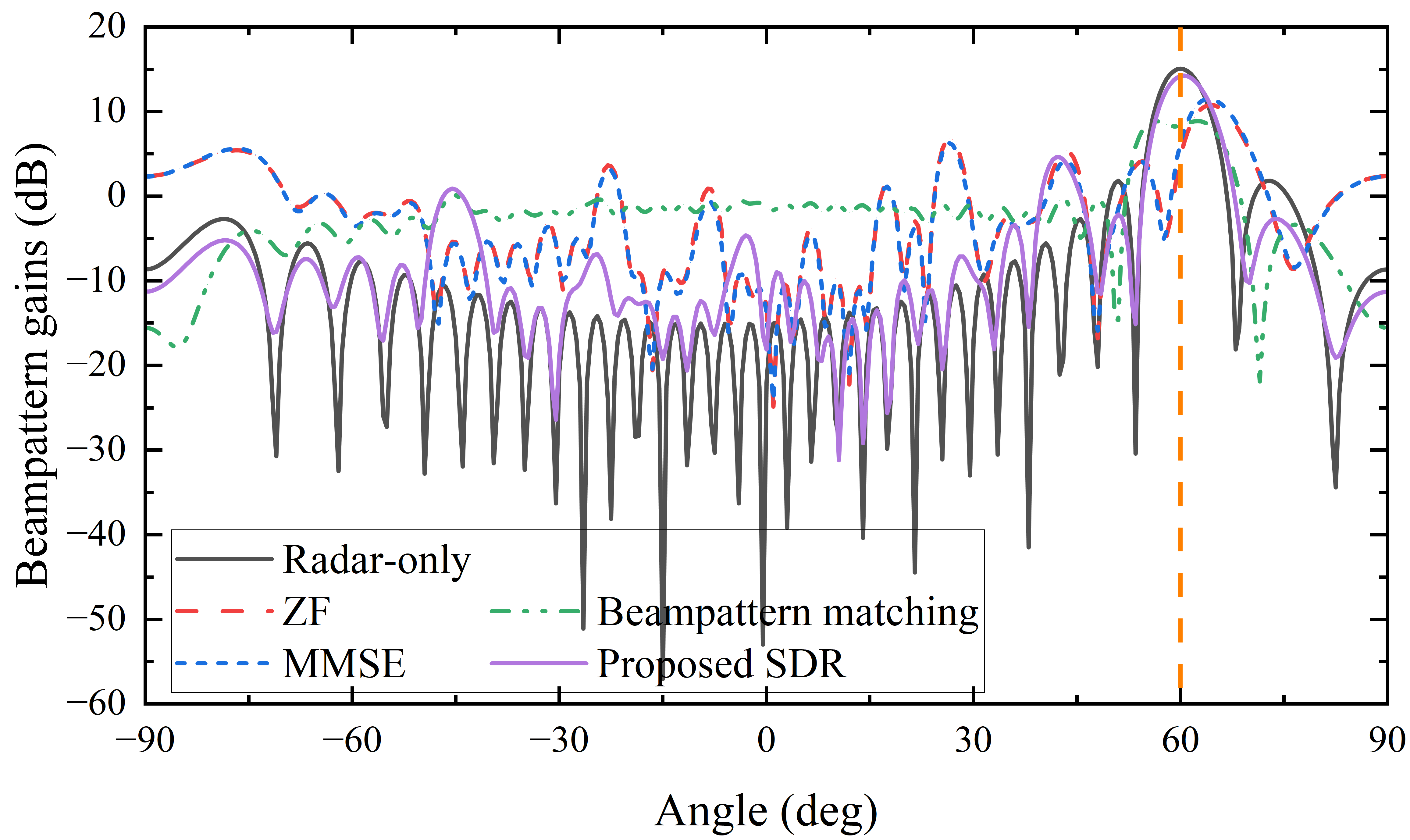}%
\label{figbeampattern1}}

\subfloat[]{\includegraphics[width=3.3in]{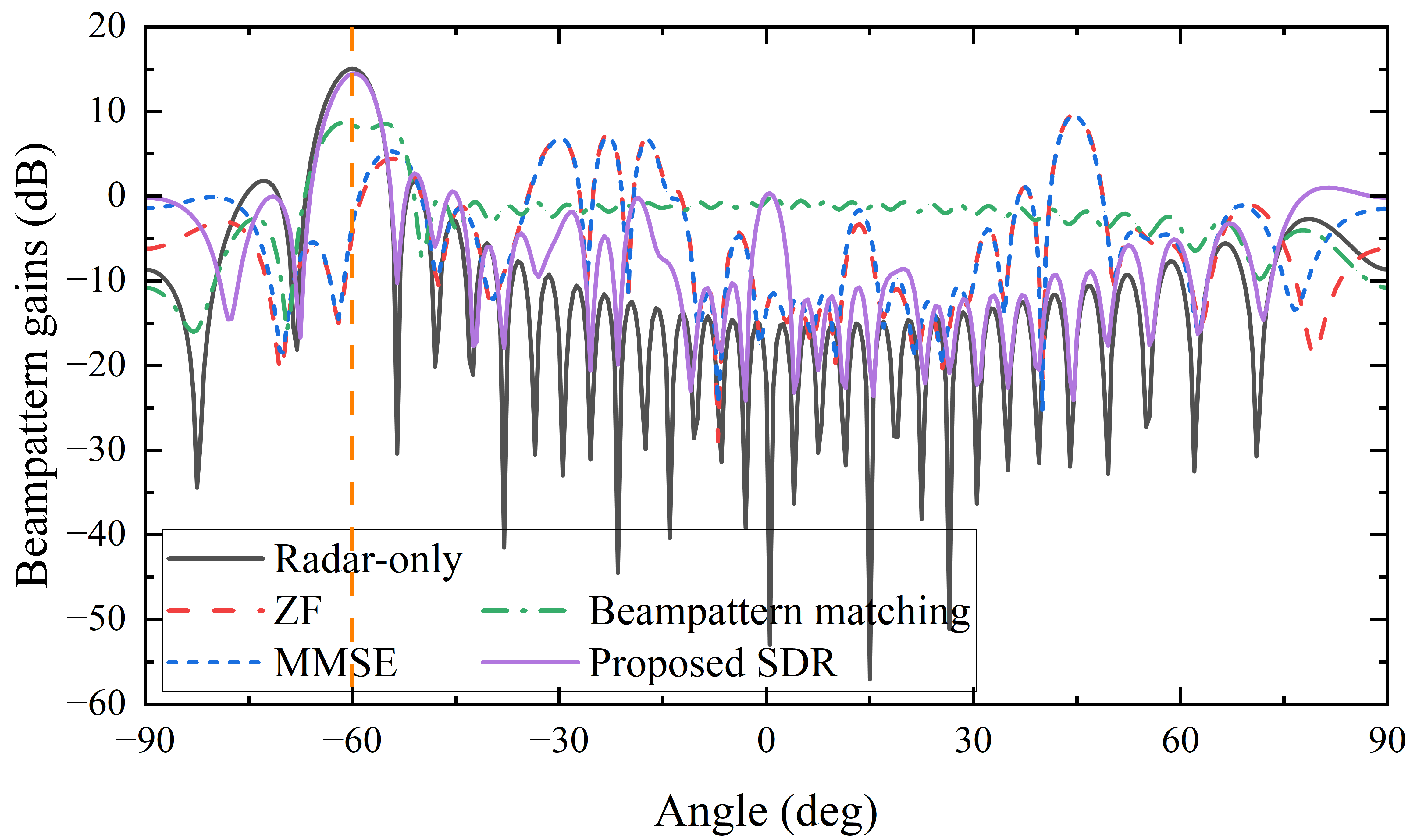}%
\label{figbeampattern2}}
\caption{Beampatterns of all the considered algorithms. (a) BS 1. (b) BS 2.}
\label{figbeampattern}
\end{figure}

We further investigate the impact of BS and TMT density on CRLB performance using Algorithm~1. 
In this setup, the number of BSs is increased   from $M=2$ to $M=4$,  
and   the number of TMTs is extended  from $N=4$ to $N=6$.
The newly added TMTs are positioned at $(0,50\sqrt{2})$m and $(0,-50\sqrt{2})$m, 
while the additional BSs are located at $(-80,80\sqrt{3})$m and $(-80,-80\sqrt{3})$m.
Crucially, since the number of CUs per BS is fixed at $K=4$, adding BSs increases the total CU population, which simultaneously augments the total system power and intensifies inter-user communication interference.
The resulting CRLB performance is depicted in Fig. \ref{figCRLBvsMvsN}. It is observed that increasing the number of BSs reduces the CRLB when the SINR threshold is low. However, under stringent SINR conditions (e.g., $\eta = 30$ dB), adding BSs counterintuitively results in an elevated CRLB. This phenomenon stems from the competing effects of BS densification. In the low SINR regime, the benefits of enhanced spatial diversity and increased total power dominate, improving localization accuracy. Conversely, in the high SINR regime, the impact of aggregated communication interference becomes the governing factor. As interference dominates, the system is forced to divert spatial degrees of freedom toward interference mitigation rather than sensing optimization, ultimately degrading localization accuracy. In contrast, increasing the number of TMTs consistently reduces the CRLB. Given the flexibility and cost-effectiveness of TMT deployment, increasing their density emerges as a robust strategy for enhancing the localization accuracy of networked ISAC systems.

\begin{figure}[t]
	\centering
	\includegraphics[width=3.3in]{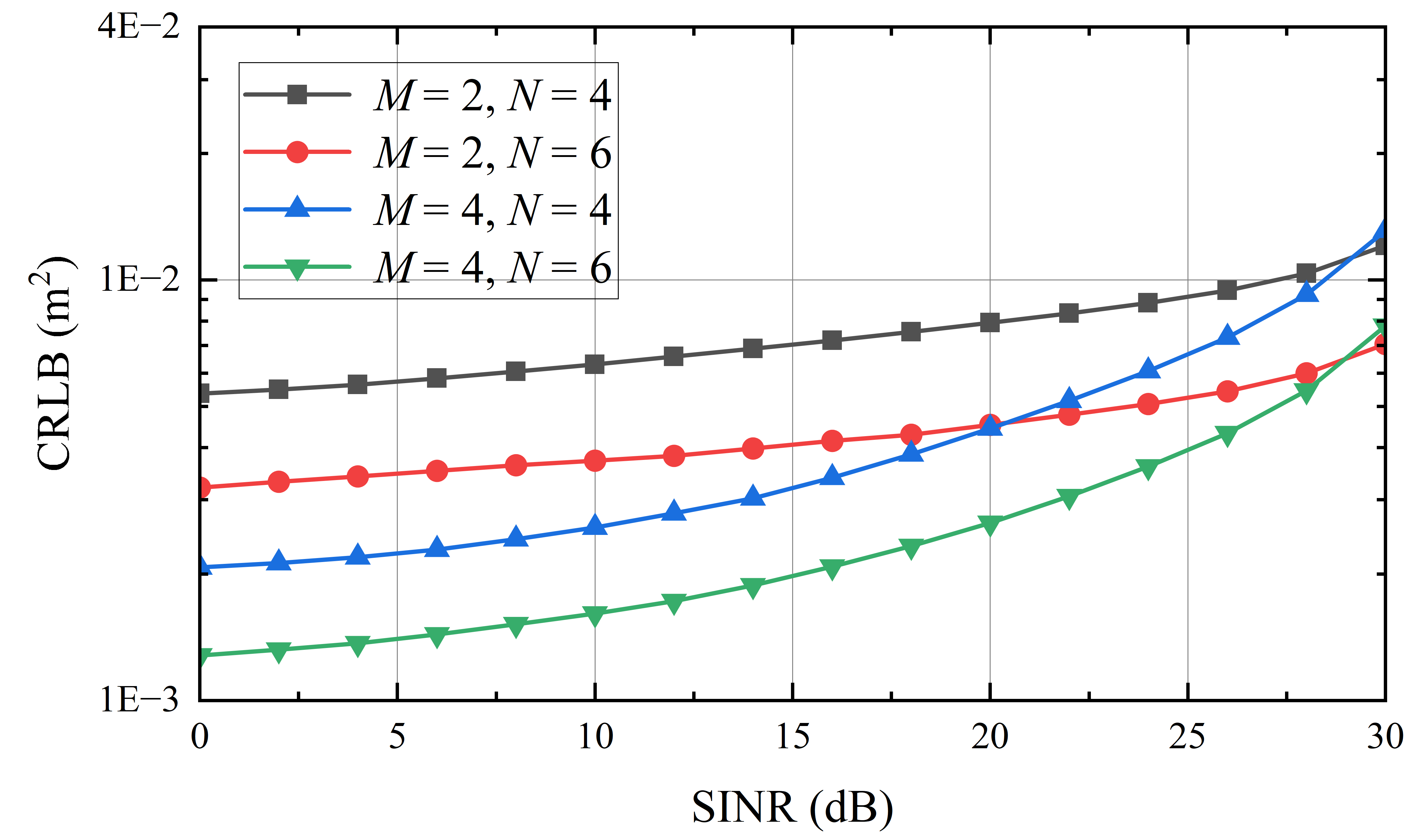}
	\caption{CRLB performance versus $M$  and $N$.}
	\label{figCRLBvsMvsN}
\end{figure}


{Subsequently, we evaluate the impact of the number of targets 
on the CRLB performance using the proposed Algorithm  \ref{alg:SCA}.
In this simulation, three scenarios are considered: a single target located at the origin, 
two targets positioned at $(0,30)\mathrm{m}$ and $(0,-30)\mathrm{m}$, and 
three targets located at $(0,30)\mathrm{m}$, $(0,-30)\mathrm{m}$, and the origin.
The   CRLB performance is illustrated in Fig. \ref{figCRLBvsU}.
It is observed that the CRLB increases with the number of targets.
This is because the presence of multiple targets forces the available 
power to be distributed among them, reducing the power allocated to each 
individual target and thereby degrading the localization accuracy.}

\begin{figure}[t]
	\centering
	\includegraphics[width=3.3in]{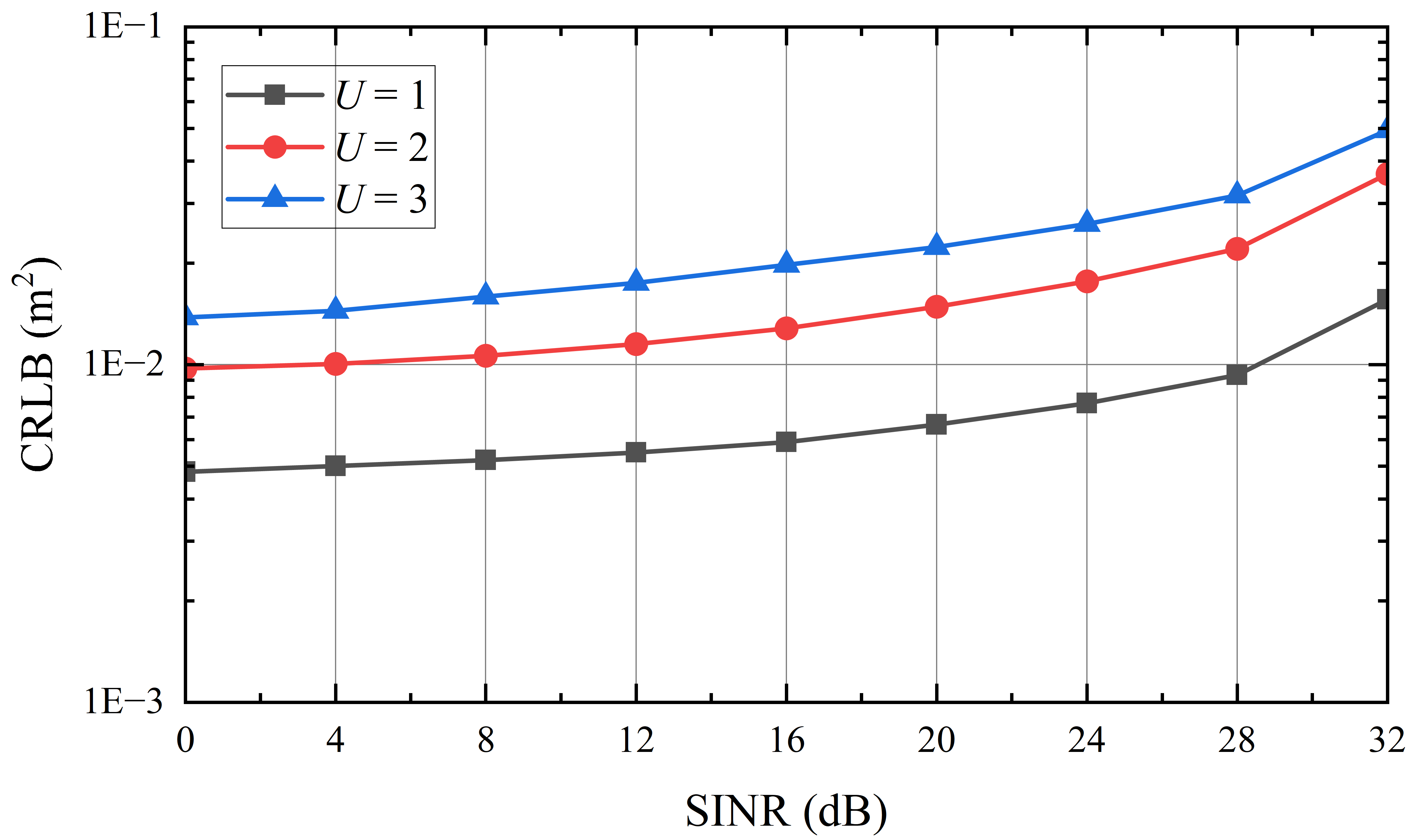}
	\caption{CRLB performance versus $U$.}
	\label{figCRLBvsU}
\end{figure}

\subsubsection{Communication-centric Beamforming Algorithms}
Next, we evaluate   the 
proposed communication-centric beamforming  algorithms
against   benchmark  algorithms.

Fig. \ref{figSINRvsEpsilon} illustrates the   maximum SINR 
achievable by each CU   
as a function of the CRLB threshold $\epsilon$. 
For the case of  $M=1$,  only the BS located at $(80,80\sqrt{3})$m is retained.
The proposed algorithms consistently outperform  the ZF and MMSE algorithms 
for both $M=1$ and $M=2$.
As   $\epsilon$ increases, 
the SINR of the proposed algorithms approaches that of the communication-only   algorithm.
This occurs because relaxing the requirement on localization accuracy 
allows the ISAC system to allocate more power towards achieving higher communication SINR.
This once again underscores the intrinsic trade-offs between communication and localization
in networked ISAC systems.
Additionally,   Algorithm \ref{alg:SCA}    
achieves   near-optimal performance compared  to Algorithm \ref{alg:BiSearch}
for the case of  $M=1$, demonstrating its ability 
to effectively balance computational complexity and performance.



\begin{figure}[t]
\centering
\subfloat[]{\includegraphics[width=3.3in]{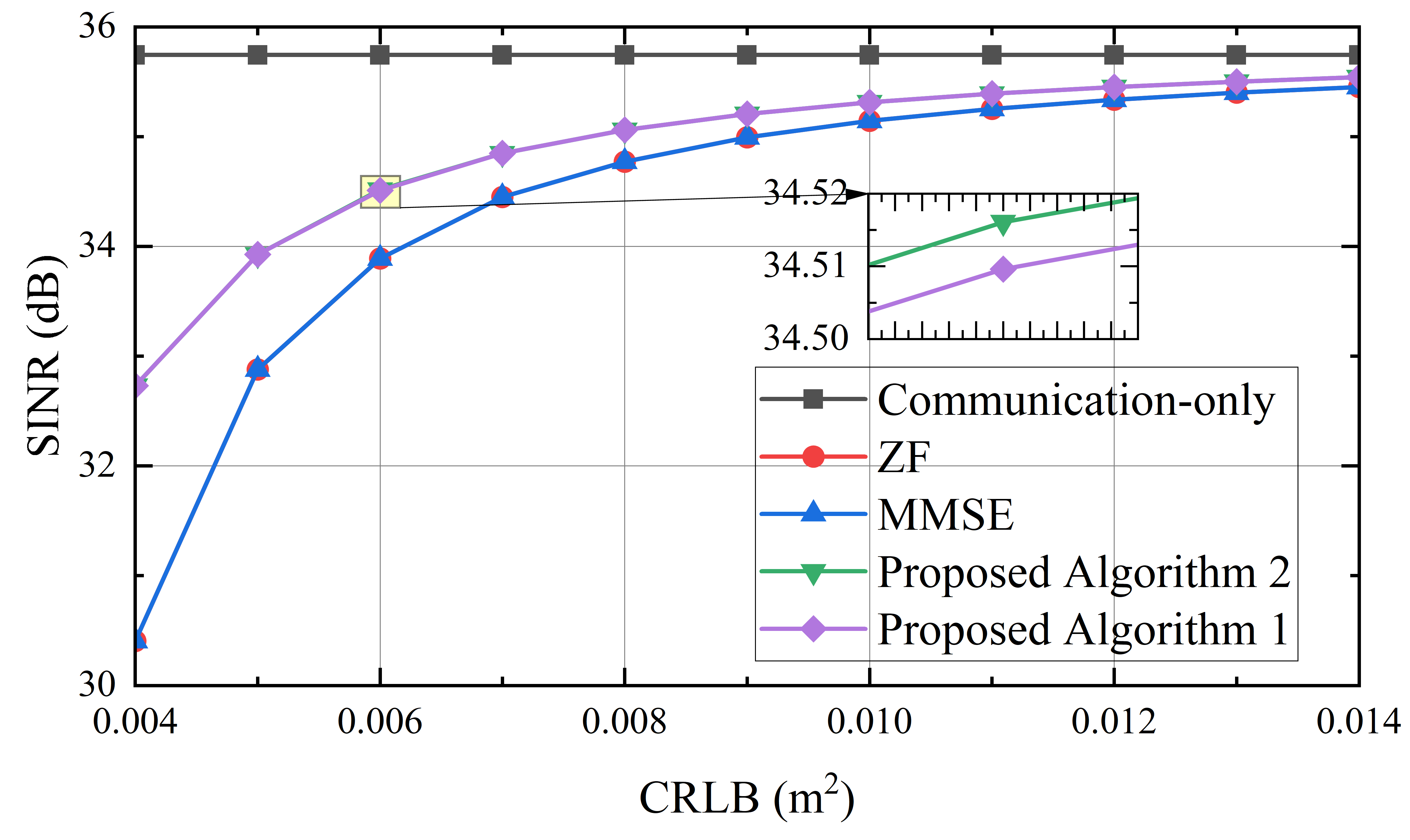}%
\label{figSINRvsEpsilon1}}

\subfloat[]{\includegraphics[width=3.3in]{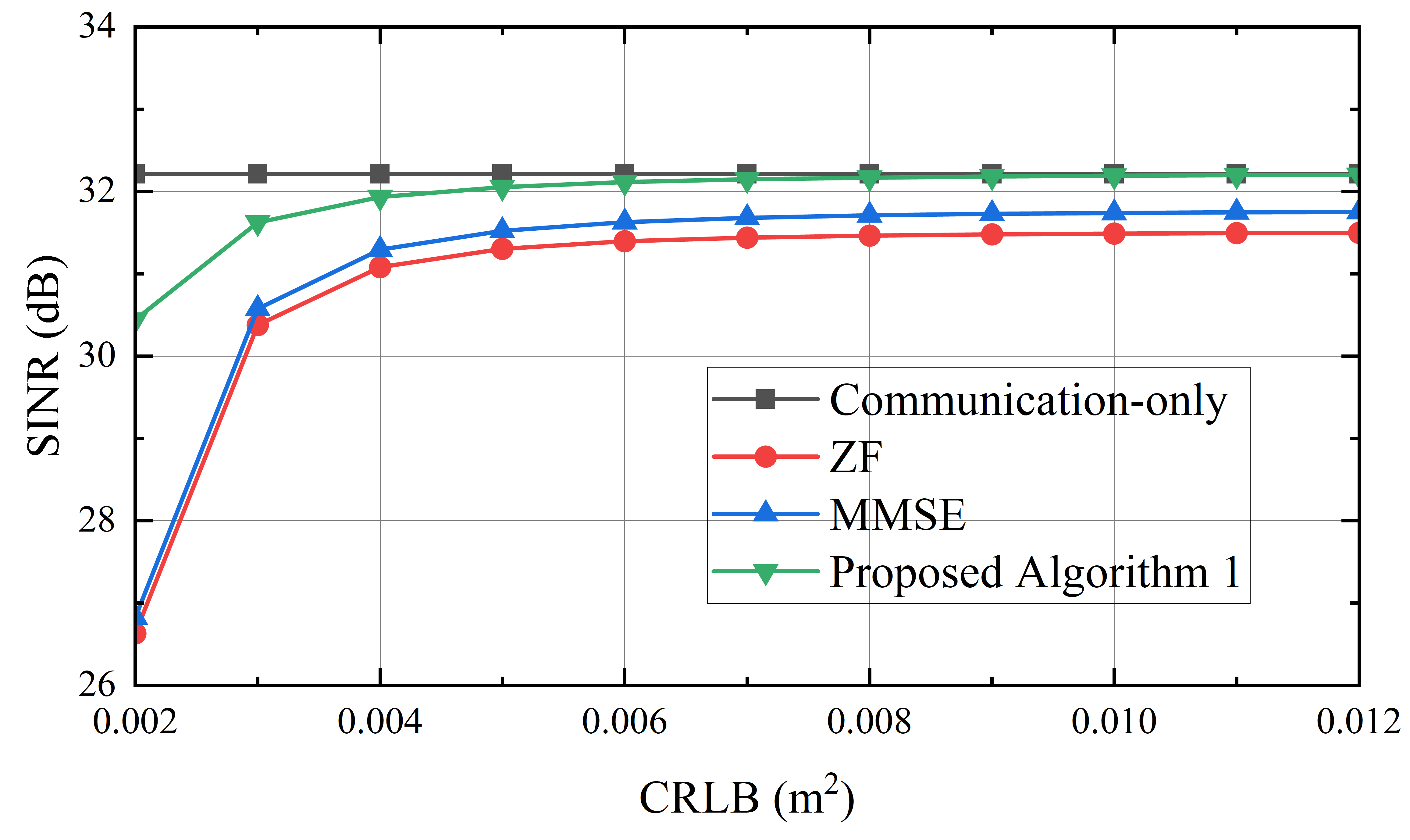}%
\label{figSINRvsEpsilon2}}
\caption{SINR performance versus $\epsilon$. (a) $M$ = 1. (b) $M$ = 2.}
\label{figSINRvsEpsilon}
\end{figure}

Subsequently, we demonstrate that the deployment of TMTs can also enhance the communication performance of networked ISAC systems. To this end, we evaluate the maximum SINR achievable by each CU using Algorithm \ref{alg:SCA} as a function of the number of BSs and TMTs, as shown in Fig. \ref{figSINRvsMvsN}. The deployment configuration mirrors that of Fig. \ref{figCRLBvsMvsN}, with the CU density per BS fixed at $K = 4$.
Notably, given the same total number of BSs and TMTs remains the same, 
the configuration $M = 2$, $N = 6$ achieves a higher SINR 
than the configuration $M = 4$, $N = 4$.
Two key factors drive this result. First, adding BSs introduces additional CUs, thereby intensifying communication interference levels which degrades SINR. Second, employing more TMTs alleviates the sensing constraint, allowing the system to allocate more power and degrees of freedom toward maximizing SINR.
These results demonstrate that TMTs not only improve localization accuracy but also boost communication performance in networked ISAC systems, indicating that prioritizing TMT deployment is a superior strategy to adding BSs for enhancing SINR.

\begin{figure}[t]
	\centering
	\includegraphics[width=3.3in]{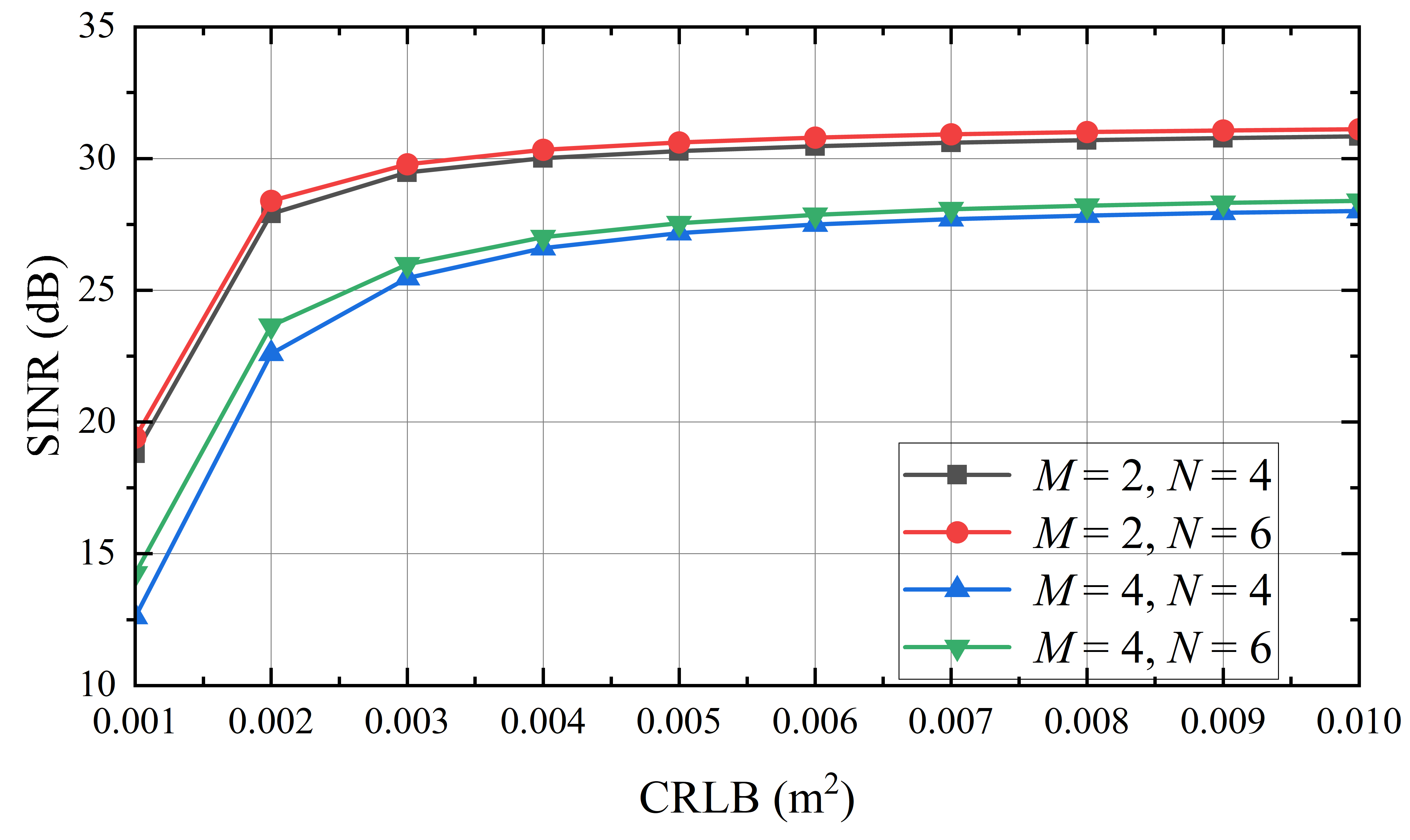}
	\caption{SINR performance versus $M$ and $N$.}
	\label{figSINRvsMvsN}
\end{figure}

{
Finally, we evaluate the impact of the number of targets on SINR performance using Algorithm \ref{alg:SCA}.
The simulation considers the same three scenarios as 
in Fig. \ref{figCRLBvsU}, with the resulting SINR performance shown in Fig. \ref{figSINRvsU}.
It is observed that SINR decreases as the number of targets increases.
This degradation occurs because supporting more targets 
necessitates allocating more power to the sensing function of the 
ISAC system to meet localization accuracy requirements, 
thereby leaving less power available for communication,
ultimately leading to reduced SINR performance.}

\begin{figure}[t]
	\centering
	\includegraphics[width=3.3in]{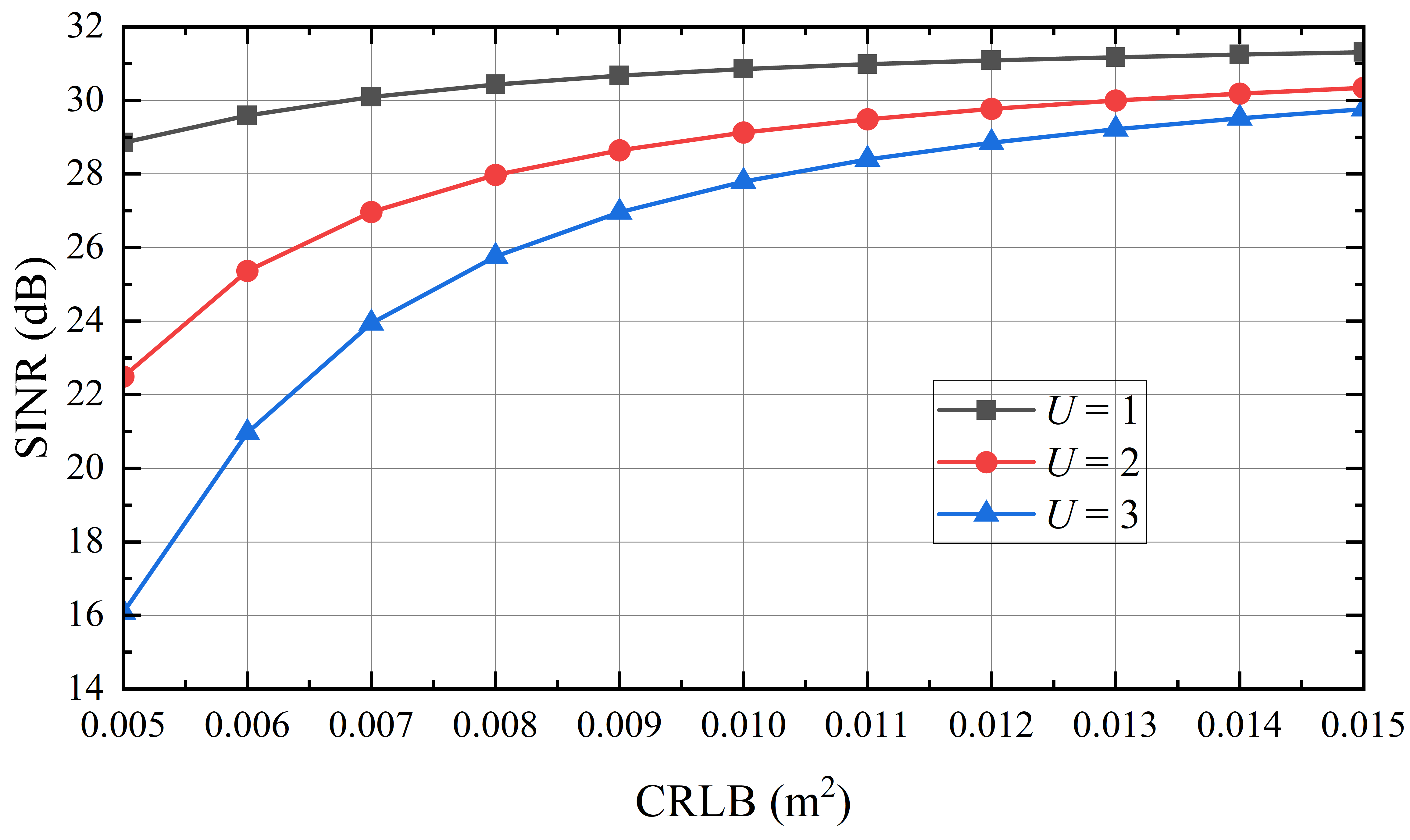}
	\caption{SINR performance versus $U$.}
	\label{figSINRvsU}
\end{figure}

\section{Conclusion}
\label{section6}
In this paper, we investigated the coordinated beamforming design for networked ISAC systems empowered by multiple TMTs. We first established the signal models for both communication and localization, deriving closed-form expressions for the communication SINR and the localization CRLB. Building upon these metrics, we formulated two nonconvex optimization problems aimed at minimizing the CRLB and maximizing the minimum SINR, respectively. To address these problems, we proposed effective algorithms leveraging SDR, bisection search, and SCA techniques. Numerical results demonstrated that the proposed algorithms achieve  satisfactory trade-offs between communication and sensing, verifying that TMTs offer advantages over BSs in terms of both communication and localization performance.

{While this work provides valuable insights into networked ISAC systems, the current results serve as benchmark designs under idealized  assumptions}. To facilitate real-world deployment, several practical challenges warrant further investigation. First, our current framework relies on assumptions of perfect channel state information, ideal clock synchronization, and interference-free sensing. In practice, synchronization errors and  interference necessitate the derivation of more accurate CRLB expressions, while channel uncertainties may significantly degrade beamforming gains. Consequently, developing robust optimization techniques that account for these imperfections is a vital direction for future work. Second, while this study assumes ideal backhaul links, real-world systems are limited by finite capacity and latency. Future research should therefore quantify the trade-offs between cooperation gains and backhaul overhead. Third, current research treats the spatial placement of TMTs and TMT-target associations as   predetermined. However, optimizing these placement and allocation strategies, particularly in multi-target scenarios, remains an open and critical problem.
Finally, as current framework does not account for data association errors and inter-target interference, extending the framework to complex multi-target environments involving these factors represents a promising direction for future exploration.

\appendices
\section{Proof of Theorem \ref{theorem1}}
\label{app:theorem1}
Let $\boldsymbol{\uppsi} = \left[\boldsymbol{\uptau}^T, \boldsymbol{\uptheta}^T, 
{\boldsymbol{\upvarepsilon}_{\mathrm{R}}}^T, 
{\boldsymbol{\upvarepsilon}_{\mathrm{I}}}^T\right]^T$ represent the parameter vector.
Here, $\boldsymbol{\uptau} = \left[\tau_{1,1}, \ldots,   \tau_{M, N}\right]^T$, 
and $\boldsymbol{\uptheta} = \left[\theta_{1}, \ldots, \theta_{M}\right]^T$. Furthermore,  
${\boldsymbol{\upvarepsilon}_{\mathrm{R}}}$ and ${\boldsymbol{\upvarepsilon}_{\mathrm{I}}}$ 
represent the real and imaginary parts  of the vector  
$\boldsymbol{\upvarepsilon} = \left[\varepsilon_{1,1}, \ldots,  \varepsilon_{M, N}\right]^T$, respectively. 
The parameters $\boldsymbol{\uptheta}$, ${\boldsymbol{\upvarepsilon}_{\mathrm{R}}}$, and ${\boldsymbol{\upvarepsilon}_{\mathrm{I}}}$ 
are considered as nuisance parameters,
which are not directly related to the ToA-based   localization.

The Fisher information  matrix 
(FIM) pertaining to the parameter vector
$\boldsymbol{\uppsi}$ is denoted by $\mathbf{J}(\boldsymbol{\uppsi})$.
Its $(l,p)$-th element is given by the 
Slepian-Bang formula  \cite{stoicaSpectralAnalysisSignals2005} 
\begin{equation}
    J(\psi_p,\psi_l)=\frac{2}{\sigma_{\mathrm{s}}^2} 
    \int_{0}^{T}\operatorname{Re}\left( \frac{\partial \boldsymbol{\upmu}^H(t)}{\partial \psi_l} 
    \frac{\partial \boldsymbol{\upmu}(t)}{\partial \psi_p}\right) \mathrm{d}t, \ \forall l, p, \label{26}
\end{equation}
where $\boldsymbol{\upmu}(t)=\left[\mu_1(t),\mu_2(t),\ldots,
\mu_N(t)\right]^T$. 

Following   \eqref{26}, we derive 
\begin{equation}
    \begin{aligned}
        & J(\tau_{m,n},\tau_{m^{\prime},n^{\prime}}) = 
        \left\{ \begin{array}[2]{ll}
            \frac{2}{\sigma_{\mathrm{s}}^2}    u_{m,n}     , & m=m^{\prime},n=n^{\prime}, \\
            0, & \text{otherwise},
        \end{array} \right.
    \end{aligned} \label{1}
\end{equation}
where 
\begin{equation}
    \begin{aligned}
        u_{m,n}  = & \left| \varepsilon_{m,n} \right|^2   \sum_{k=1}^{K} \mathbf{a}^H\left(\theta_m\right)  
        \mathbf{f}_{m, k} \mathbf{f}^{H}_{m, k}
        \mathbf{a}\left(\theta_m\right) \\ & \times   \int_{0}^{T} \left|
        \dot{s}_{m,k}(t-\tau_{m,n}) \right|^2 \mathrm{d}t
    \end{aligned}
\end{equation}
and $\dot{s}_{m,k}(t-\tau_{m,n})=\frac{\partial s_{m,k}(t-\tau_{m,n})}{\partial \tau_{m,n}}$.
Leveraging the {Parseval's} theorem, we can obtain 
\cite{lehmannHighResolutionCapabilities2006}  
\begin{equation}
    \hspace{-0.3cm} \int_{0}^{T} \left|\dot{s}_{m,k}(t-\tau_{m,n}) \right|^2 \mathrm{d} t= 4\pi^2 L
    \int_{-\infty}^{\infty} f^2 \left|G(f)\right|^2 \mathrm{d}f.
\end{equation}
Exploiting the effective bandwidth $\beta$, we have 
$u_{m,n}  =   4\pi^2 T \beta^2 \left| \varepsilon_{m,n} \right|^2   
        \mathbf{a}^H\left(\theta_m\right) 
        \left( 
        \sum_{k=1}^{K}\mathbf{f}_{m,k} \mathbf{f}_{m,k}^{H} \right)  \mathbf{a}\left(\theta_m\right)$. 
Similarly, we establish the following equations
\begin{subequations}
    \begin{align}
        J(\tau_{m,n},\theta_{m^{\prime}}) &= 0, \  \forall m, n,  m^{\prime},  \label{22} \\
        J(\tau_{m,n},\varepsilon_{m^{\prime},n^{\prime},\mathrm{R}}) &= 0, \ \forall m, n, m^{\prime}, n^{\prime},  \label{3} \\
        J(\tau_{m,n},\varepsilon_{m^{\prime},n^{\prime},\mathrm{I}}) &= 0, \ \forall m, n, m^{\prime}, n^{\prime}, \label{44}
    \end{align}
\end{subequations}
where $\varepsilon_{m^{\prime},n^{\prime},\mathrm{R}}$ and
$\varepsilon_{m^{\prime},n^{\prime},\mathrm{I}}$ represent the real 
and imaginary part of $\varepsilon_{m^{\prime},n^{\prime}}$, respectively.
From   \eqref{22} to   \eqref{44}, we have used 
the fact that \cite{lehmannHighResolutionCapabilities2006} 
\begin{equation}
    \begin{aligned}
        \int_{0}^{T} & \dot{s}^{*}_{m,k}(t-\tau_{m,n}) s_{m,k}(t-\tau_{m,n}) \mathrm{d}t \\ 
        & = -\jmath 2\pi L \int_{-\infty}^{\infty} f \left|G(f)\right|^2 \mathrm{d}f = 0.
    \end{aligned}
\end{equation}

It is clear that the FIM $\mathbf{J}(\boldsymbol{\uppsi})$
is a diagonal matrix  as denoted by
\begin{equation}
    \label{Jpsi}
    \mathbf{J}(\boldsymbol{\uppsi})  =  
    \begin{bmatrix}
        \mathbf{Z}  &  \mathbf{0}_{MN \times (2MN+M)} \\
        \mathbf{0}_{  (2MN+M) \times MN}  &   {\boldsymbol{\Omega}} \\
    \end{bmatrix},
\end{equation}
where $\mathbf{Z} \in \mathbb{C}^{MN \times MN}$ is described in $\eqref{Z}$ and $\eqref{J_ele}$, while
$\mathbf{\boldsymbol{\Omega}} \in \mathbb{C}^{(2MN+M) \times (2MN+M)}$ is the Fisher information submatrix
with respect to the nuisance parameters.

By the chain rule \cite{xuOptimalSensorPlacement2019}, the FIM for the target location $(x,y)$ is given by
$\mathbf{C} = \boldsymbol{\Lambda} \mathbf{Z} \boldsymbol{\Lambda}^T$, 
where the elements of $\boldsymbol{\Lambda}$ are computed from \eqref{tau}.
Since the diagonal elements of the inverse FIM $\mathbf{C}^{-1}$ represent the CRLB for the target coordinates \cite{kayFundamentalsStatisticalSignal2013a}, the total localization error bound can be expressed as the trace of $\mathbf{C}^{-1}$, as given in $\eqref{CRLBxy}$.

\section{Proof of Lemma \ref{lemma2}}
\label{app:lemma1}
We first prove that $\operatorname{tr}\Big(\Big(\mathbf{\Lambda} \hat{\mathbf{Z}} 
\mathbf{\Lambda}^T\Big)^{-1}\Big)$
is a  monotonically decreasing function of $q_m, \ \forall m$.
To this end, we calculate the derivative  
\begin{equation}
    \hspace{-0.2cm}
    \begin{aligned}
        & \frac{\partial}{\partial q_m} \operatorname{tr}\Big(\Big(\mathbf{\Lambda} \hat{\mathbf{Z}}
        \mathbf{\Lambda}^T\Big)^{-1}\Big)  \\
        & = -\operatorname{tr}\Big(\Big(\mathbf{\Lambda} \hat{\mathbf{Z}} \mathbf{\Lambda}^T\Big)^{-1} 
        \hat{\mathbf{\Lambda}}_m \hat{\mathbf{Z}}_m \hat{\mathbf{\Lambda}}_m^T
        \left(\mathbf{\Lambda} \hat{\mathbf{Z}} \mathbf{\Lambda}^T\right)^{-1} \Big) \\
        & < 0, \ \forall m, \label{ineq1}
    \end{aligned}
\end{equation}
where $\hat{\mathbf{\Lambda}}_m$ is formed by the corresponding columns of $\mathbf{\Lambda}$,
and the last inequality holds because that $\left(\mathbf{\Lambda} \hat{\mathbf{Z}} \mathbf{\Lambda}^T\right)^{-1} 
\hat{\mathbf{\Lambda}}_m \hat{\mathbf{Z}}_m 
\hat{\mathbf{\Lambda}}_m^T\left(\mathbf{\Lambda} \hat{\mathbf{Z}} \mathbf{\Lambda}^T\right)^{-1}$
is a positive semidefinite matrix and is not a zero matrix.
From \eqref{ineq1}, it implies that 
$\operatorname{tr}\Big(\Big(\mathbf{\Lambda} \hat{\mathbf{Z}} \mathbf{\Lambda}^T\Big)^{-1}\Big)$
is a  monotonically decreasing function of $q_m, \ \forall m$.
If  \eqref{qqq} is not blinding at the optimum,
then we can strictly increase $\left\{ q_m \right\}$ which reduces the objective \eqref{ob}.
Therefore,  constraint  \eqref{qqq} must be active at the optimum,
implying that    problem   \eqref{eq:problem_q} is equivalent to   original problem   \eqref{eq:problem}.

\section{Proof of Theorem \ref{theorem2}}
\label{app:theorem2}
We analyze under the assumption that  problem \eqref{eq:problem_eq} is feasible,
as commonly adopted in the literature \cite{heFullDuplexCommunicationISAC2023a}.
It can be verified that
problem \eqref{eq:problem_eq} 
is a convex optimization problem and the  Slater's condition holds,
ensuring that  strong duality holds \cite{BoydConvexOptimization}. 
Subsequently, we complete the proof of Theorem 2
by analyzing the KKT conditions of   problem  \eqref{eq:problem_eq}.

The dual variables for problem \eqref{eq:problem_eq} are defined as 
$\left\{\lambda_{m}\right\}  \geq 0$ for   \eqref{SDPb},
$\left\{ \gamma_{m,k}\right\} \geq 0$
for   \eqref{SDPc}, $\left\{ \mu_m\right\} \geq 0$ 
 for  \eqref{SDPd}, $\left\{ \iota_m \right\} \geq 0$ for 
 \eqref{q_nonge2},
and $\left\{ \mathbf{S}_{m,k}\right\} \succeq \mathbf{0}$ 
for   \eqref{SDPe}. 
Then, the Lagrangian function of problem \eqref{eq:problem_eq} is given by \eqref{Lag} at the bottom of next page.
\begin{figure*}[!b]
\begin{equation}
    \label{Lag}
    \begin{aligned}
        & \mathcal{L}\left(\left\{\mathbf{F}_{m,k}\right\}, \left\{ q_m \right\},
        \left\{\lambda_{m}\right\}, \left\{\gamma_{m,k}\right\},
        \left\{\mu_m\right\}, \left\{\iota_m\right\}, \left\{\mathbf{S}_{m,k}\right\}\right)  =    \mathrm{tr}\Big( \Big(\boldsymbol{\Lambda} \hat{\mathbf{Z}} 
        \boldsymbol{\Lambda}^T\Big)^{-1} \Big)
        + \sum_{m=1}^{M} \lambda_m \Big( \mathrm{tr} \Big( \sum_{k=1}^{K} 
        \mathbf{F}_{m,k} \Big) - P_m \Big)  \\ & - \sum_{m=1}^M \sum_{k=1}^{K} \gamma_{m,k} \Big( 
        \mathbf{h}_{m,m,k}^{H}\mathbf{F}_{m,k}\mathbf{h}_{m,m,k}    - \eta \Big( \sum_{(i,j) \neq (m,k)}    \mathbf{h}_{i, m, k}^{H}
        \mathbf{F}_{i, j}\mathbf{h}_{i, m, k} +
        \sigma_{\mathrm{n}}^2 \Big) \Big) - \sum_{m=1}^{M} \sum_{k=1}^{K} \mathrm{tr} \left( \mathbf{S}_{m,k} \mathbf{F}_{m,k} \right)  \\
        & + \sum_{m=1}^{M} \mu_m \Big(q_m-\mathbf{a}^H\left(\theta_m\right) 
        \Big(\sum_{k=1}^{K}\mathbf{F}_{m,k} \Big)\mathbf{a}\left(\theta_m\right)\Big)  - \sum_{m=1}^{M} \iota_m q_m.  
    \end{aligned}  
\end{equation}
\end{figure*}
Assume that the Lagrangian function reaches its optimum at
$\left\{ \mathbf{F}_{m,k}^{\star}\right\}$, 
$\left\{ \lambda_{m}^{\star}\right\} $, $\left\{ \gamma_{m,k}^\star\right\}$,
$\left\{ \mu_m^{\star} \right\}$, $\left\{ \iota_m^{\star} \right\}$, and
$\left\{ \mathbf{S}_{m,k}^{\star}\right\}$.
The Lagrangian function is denoted by $\mathcal{L}$ for 
convenience in the following discussion, unless otherwise specified.
According to the KKT 
conditions \cite{BoydConvexOptimization},
we have
\begin{subequations}
    \begin{align}
    & \ \ \ \begin{aligned}
		\frac{\partial \mathcal{L}}{\partial \mathbf{F}_{m,k}}  = \ &
        \lambda_m^{\star} \mathbf{I}_{N_{\mathrm{t}}}
        - \gamma_{m,k}^{\star} \mathbf{h}_{m,m,k} \mathbf{h}_{m,m,k}^H    \\
        &+     \sum_{(i,j) \neq (m,k)}  \gamma_{i,j}^{\star} \eta \mathbf{h}_{m,i,j} \mathbf{h}_{m,i,j}^H 
        - \mathbf{S}_{m,k}^{\star}   \\ 
        &- \mu_m^{\star} \mathbf{a}\left(\theta_m\right) \mathbf{a}^H\left(\theta_m\right) 
        = \mathbf{0}, \ \forall m,k, 
		\end{aligned} \label{566} \\
	&  \mathbf{S}_{m,k}^{\star} \mathbf{F}_{m,k}^{\star} = \mathbf{0}, \ \forall m,k. \label{567}
    \end{align}        
\end{subequations}
From \eqref{566},
we have
\begin{equation}
    \begin{aligned}
        \mathbf{S}_{m,k}^{\star}   = \ & \lambda^{\star}_m \mathbf{I}_{N_{\mathrm{t}}} +   
        \mathbf{D}^{\star}_{m}- \mu_m^{\star} \mathbf{a}(\theta_m) \mathbf{a}^H(\theta_m) \\
		&-  \gamma_{m,k}^{\star} \left( 1+\eta \right)
        \mathbf{h}_{m,m,k} \mathbf{h}_{m,m,k}^H \succeq \mathbf{0}, \ \forall m,k, \label{40} 
    \end{aligned}  
\end{equation}
where 
$\mathbf{D}_{m}^{\star} =   \sum_{i,j} \gamma_{i,j}^{\star} \eta
            \mathbf{h}_{m,i,j} \mathbf{h}_{m,i,j}^H, \ \forall m.$
From  \eqref{567}, 
we have $\mathrm{rank}\left(\mathbf{S}_{m,k}^{\star}\right) + 
\mathrm{rank}\left(\mathbf{F}_{m,k}^{\star}\right) \leq N_{\mathrm{t}}, \ \forall m,k.$

We now show that $\lambda_m^{\star} > 0, \ \forall m$ when  $\mathbf{a}\left(\theta_m\right) \notin 
\mathrm{span}\left( \bigcup_{i,j} \mathbf{h}_{m,i,j} \right), \ \forall m$. 
To this end,
we first prove that the  power constraint in  \eqref{SDPb}
is always active at the optimum.
We assume that the power constraint in \eqref{SDPb} is not binding at the optimum
for the $m$-th BS, i.e., $\mathrm{tr}\left(\sum_{k=1}^{K} \mathbf{F}_{m,k}^{\star}\right) < P$.
The eigenvalue decomposition of $\mathbf{F}_{m,k}^{\star}$ is   given by
$\mathbf{F}_{m,k}^{\star}= \sum_{r=1}^{R_{m,k}} \alpha_{m,k}^{r}
    \mathbf{w}_{m,k}^{r} \left(\mathbf{w}_{m,k}^{r}\right)^H$,
where $\mathbf{w}_{m,k}^{r}$ is the $r$-th eigenvector with 
$\alpha_{m,k}^{r}$ as the corresponding eigenvalue, and $R_{m,k}$ represents  the   rank
of $\mathbf{F}_{m,k}^{\star}$.
The given condition  implies 
$\sum_{l=1}^{a_{m}}  
\mathbf{\Pi}_{\mathbf{e} _{{m}}^{l}}\mathbf{a}(\theta_m)
\neq \mathbf{a}(\theta_m), \ \forall m$,
where $\left\{\mathbf{e} _{{m}}^{l}\right\}$
is a set of
orthogonal basis vectors  for $\mathrm{span}
\left( \bigcup_{i,j} \mathbf{h}_{m,i,j} \right)$, 
and $a_{m}$ is the dimension of $\mathrm{span}
\left( \bigcup_{i,j} \mathbf{h}_{m,i,j} \right)$.
Define $\mathbf{u} _{m} = \left(\mathbf{I}_{N_{\mathrm{t}}}-\sum_{l=1}^{a_{m}}  
    \mathbf{\Pi}_{\mathbf{e} _{{m}}^{l}}\right)\mathbf{a}(\theta_m)$.
This ensures that $\mathbf{u} _{m}$ lies in the null space of
$\mathrm{span}
\left( \bigcup_{i,j} \mathbf{h}_{m,i,j} \right)$.
Next, we can choose  subscript $k$ and  superscript $r$ arbitrarily and replace
$\mathbf{w}_{m,k}^{r}$ as  
\begin{equation}
    \begin{aligned}
        \mathbf{w}_{m,k}^{r, \mathrm{new}} = \mathbf{w}_{m,k}^{r} + \delta_{m} \mathbf{u} _{m}
        e^{ \jmath \arg \left( \mathbf{a}^H(\theta_m) \mathbf{w}_{m,k}^{r} \right)},
    \end{aligned} \label{replacement}
\end{equation}
where $\delta_{m}$ is a positive scalar.
Then, we have 
\begin{subequations}
    \begin{align}
        \left| \mathbf{a}^H(\theta_m)\mathbf{w}_{m,k}^{r, \mathrm{new}}\right|^2 
        = \ &\left( \left| \mathbf{a}^H(\theta_m)\mathbf{w}_{m,k}^{r} \right| + \delta_m \mathbf{a}^H(\theta_m) 
        \mathbf{u}_m\right)^2  \nonumber \\ 
        > \ &\left| \mathbf{a}^H(\theta_m)\mathbf{w}_{m,k}^{r}\right|^2,  \label{qnew} \\
        \left| \mathbf{h}_{m,i,j}^H \mathbf{w}_{m,k}^{r, \mathrm{new}} \right|^2  = \ &
        \left| \mathbf{h}_{m,i,j}^H\mathbf{w}_{m,k}^{r}\right|^2, \ \forall i,j. \label{hnew}
    \end{align} \label{qhnew}%
\end{subequations}
The equation in \eqref{hnew} holds for the reason that the equation
$\mathbf{h}_{m,i,j}^H \mathbf{u}_m = 0, \ \forall i,j$ holds.
Equation \eqref{qhnew} implies that   the replacement in \eqref{replacement} exclusively boosts the power in the target direction, while maintaining the power directed towards each CU unaltered. Since a higher target illumination power yields a lower CRLB (as indicated in \eqref{ineq1}) and the constant user power guarantees unchanged communication SINR, this implies that the replacement in \eqref{replacement} can effectively reduce the CRLB by employing a larger $\delta_m$ without compromising the communication  performance.


{
Subsequently, we demonstrate that the $m$-th BS can always scale up $\delta_m$ to minimize the CRLB until the power constraint in \eqref{SDPb} becomes active.
Let $P_{m,k}^{r,\mathrm{new}} $ denote the transmit power associated with the updated   vector,
which is given by 
\begin{equation}
    \label{eq:power_expansion}
	\begin{aligned}
		P_{m,k}^{r, \mathrm{new}} = \ &  
		\left(\mathbf{w}_{m,k}^{r, \mathrm{new}}\right)^H \mathbf{w}_{m,k}^{r, \mathrm{new}}\\
		= \ & P_{m,k}^{r} + \delta_{m}^2 \left\| \mathbf{u}_m \right\|^2 + 2 \delta_{m} w_{m,k},
	\end{aligned} 
\end{equation}
where $P_{m,k}^{r} = \left(\mathbf{w}_{m,k}^{r}\right)^H \mathbf{w}_{m,k}^{r}$ is original power
and $ w_{m,k} = \mathrm{Re} 
            \left(   \mathbf{u}_m^H \mathbf{w}_{m,k}^{r} e^{ -\jmath \arg 
		\left( \mathbf{a}^H(\theta_m) \mathbf{w}_{m,k}^{r} \right)} \right)$.
We observe that $P_{m,k}^{r, \mathrm{new}}$ is a convex quadratic function with respect to $\delta_m$. This implies that there always exists a sufficiently large $\delta_m > 0$ such that the transmit power reaches the maximum budget defined in \eqref{SDPb}.
Since the CRLB is monotonically decreasing in $\delta_m$, the optimal strategy dictates increasing $\delta_m$ until the power constraint becomes active. This outcome contradicts the initial assumption of an inactive constraint. Consequently, we conclude that the power constraint in \eqref{SDPb} must be active at optimality whenever $\mathbf{a}(\theta_m) \notin \mathrm{span}( \bigcup_{i,j} \mathbf{h}_{m,i,j} ), \ \forall m$.

}

{
We then show that $\lambda_m^{\star} > 0, \ \forall m$ by sensitivity analysis \cite{BoydConvexOptimization}.
First, we relax the power constraints in \eqref{SDPb}  to
$\mathrm{tr}\left(\sum_{k=1}^{K} \mathbf{F}_{m,k}\right) \leq P + \varepsilon_m, \ \forall m$,
where $\varepsilon_m$ is a small positive scalar. 
This yields a perturbed problem, whose optimal value 
we denote by $V\left(\left\{\varepsilon_m\right\}\right)$. 
The unperturbed problem in \eqref{eq:problem_eq} corresponds to $V(\left\{0\right\})$.
By strong duality, for the unperturbed problem \eqref{eq:problem_eq},
we have 
$V\left( \left\{ 0 \right\} \right)   \leq      \mathcal{L}\Big(\left\{\mathbf{F}_{m,k}\right\}, 
        \left\{ q_m \right\}, 
        \left\{\lambda_{m}^{\star}\right\}, \left\{\gamma_{m,k}^{\star}\right\},
        \left\{\mu_m^{\star}\right\}, \left\{ \iota_m^{\star} \right\}, \left\{\mathbf{S}_{m,k}^{\star}\right\}\Big)
$ \cite{BoydConvexOptimization}.
For any feasible solution of the perturbed problem, this gives
$V\left( \left\{ 0 \right\} \right) \leq   \mathrm{tr}\Big( \Big(\boldsymbol{\Lambda} \hat{\mathbf{Z}} 
\boldsymbol{\Lambda}^T\Big)^{-1} \Big) + \sum_{m=1}^{M} \lambda_m^{\star} \varepsilon_m.$
Therefore, the optimal value of the perturbed problem satisfies
\begin{equation}
	\begin{aligned}
		V( \left\{ \varepsilon_m \right\} ) \geq  \ & V( \left\{ 0 \right\} ) - 
        \sum_{m=1}^{M} \lambda_m^{\star} \varepsilon_m.
		\label{Vbound}
	\end{aligned}
\end{equation}
From the earlier analysis, increasing $P$ by any small $\varepsilon_m>0$ strictly reduces the optimal value, 
i.e.,  $V( \left\{ \varepsilon_m \right\} ) < V( \left\{ 0 \right\} )$.
Combining this with \eqref{Vbound}, we must have $\lambda_m^{\star} > 0, \ \forall m$.
}

{
Subsequently, we show that the rank of 
$\mathbf{S}_{m,k}^{\star}$ is always ${N_{\mathrm{t}}}-1, \ \forall m,k$.
For each $m$, if $\mu_m^{\star} = 0$, the problem reduces to the conventional 
communication scenario, in which it is evident 
that $\mathrm{rank}(\mathbf{S}_{m,k}^{\star}) = {N_{\mathrm{t}}}-1, 
\ \forall m, k$  \cite{heFullDuplexCommunicationISAC2023a}.
Therefore, we focus on the case $\mu_m^{\star} > 0, \ \forall m$ and analyze $\gamma_{m,k}^{\star}$.
For each $(m,k)$ pair, there are two possibilities: $\gamma_{m,k}^{\star} = 0$ or $\gamma_{m,k}^{\star} > 0$.
The two cases are examined in detail as follows.
}

{
\noindent\textbf{Case I:}  $\gamma_{m,k}^{\star} = 0$.
In this case, we have
$\mathbf{S}_{m,k}^{\star} = \lambda_m^{\star} \mathbf{I}_{N_{\mathrm{t}}} + \mathbf{D}_{m}^{\star}
		- \mu_m^{\star} \mathbf{a}(\theta_m) \mathbf{a}^H(\theta_m)$.
It is easy to show that $\mathrm{rank}\left(\mathbf{S}_{m,k}^{\star}\right) = N_{\mathrm{t}}-1$
\cite{heFullDuplexCommunicationISAC2023a}.
}

{
\noindent\textbf{Case II:} $\gamma_{m,k}^{\star} > 0$. 
Let
$\mathbf{B}_{m}^{\star} = \lambda_m^{\star} \mathbf{I}_{N_{\mathrm{t}}} + 
		\mathbf{D}_{m}^{\star} - \mu_m^{\star} \mathbf{a}(\theta_m) \mathbf{a}^H(\theta_m)
		\succeq \mathbf{0},$
where the positive semidefinite condition holds because
of its violation will lead to the violation of the positive semidefinite condition of \eqref{40}.
Given the optimal 
dual variables $\left\{\lambda_m^{\star}\right\}$, $\left\{\gamma_{m,k}^{\star}\right\}$, 
$\left\{\mu_m^{\star}\right\}$, 
and $\left\{ \iota_m^{\star} \right\}$,
the optimal value of problem \eqref{eq:problem_eq}
can, by   strong duality,  be obtained by solving the following problem
\begin{equation}
	\begin{aligned}
		\min_{ \left\{\mathbf{F}_{m,k} \succeq \mathbf{0}, \ q_m \right\} } 
		& \Delta_{m,k} +  \mathrm{tr}\left( \mathbf{F}_{m,k}  \left(
        \mathbf{B}_m^{\star} \right. \right. \\
        & \left. \left. 
		- \gamma_{m,k}^{\star} \left( 1+\eta \right)
        \mathbf{h}_{m,m,k} \mathbf{h}_{m,m,k}^H\right) \right), 
	\end{aligned} \label{eq:problem_eq2}
\end{equation}
where $\Delta_{m,k}$ is the term that does not depend on $\mathbf{F}_{m,k}$.
Subsequently, we show that $\mathbf{B}_{m}^{\star} \succ \mathbf{0}$ 
with probability one by contradiction.
Assume that   $\mathbf{B}_{m}^{\star}$ is not positive definite,
then we can find a non-zero vector $\mathbf{w}_{m}$ such that
$
	\mathbf{B}_{m}^{\star} \mathbf{w}_{m} = \mathbf{0}.
$
Unfolding   and rearranging this, we have 
$\mathbf{w}_{m} = c_m \left( \lambda_m^{\star} \mathbf{I}_{N_{\mathrm{t}}} +
		\mathbf{D}_{m}^{\star} \right)^{-1} \mathbf{a}(\theta_m),$
where 
$c_m = c_m \mu_m^{\star} \mathbf{a}^H(\theta_m) \left( \lambda_m^{\star} \mathbf{I}_{N_{\mathrm{t}}} +
			\mathbf{D}_{m}^{\star} \right)^{-1} \mathbf{a}(\theta_m).$
Therefore, only when
\begin{equation}
	\begin{aligned}
		\frac{1}{\mu_m^{\star}} = \ & \mathbf{a}^H(\theta_m) \left( \lambda_m^{\star} \mathbf{I}_{N_{\mathrm{t}}} +
			\mathbf{D}_{m}^{\star} \right)^{-1} \mathbf{a}(\theta_m),
	\end{aligned} \label{eq:condition}
\end{equation}
holds, we have 
$
	\mathbf{w}_m   
			\in  \mathrm{span} \Big(  \left( \lambda_m^{\star} \mathbf{I}_{N_{\mathrm{t}}} +
				\mathbf{D}_{m}^{\star} \right)^{-1} \mathbf{a}(\theta_m)   \Big).
$
Otherwise, $\mathbf{w}_m =   \mathbf{0}$,
which means that $\mathbf{B}_{m}^{\star} \succ \mathbf{0}$.
Even the condition in \eqref{eq:condition} holds,
due to the independence of the channel, the  equation 
$\mathbf{h}_{m,m,k}^H \left( \lambda_m^{\star} \mathbf{I}_{N_{\mathrm{t}}} +
		\mathbf{D}_{m}^{\star} \right)^{-1} \mathbf{a}(\theta_m) \neq 0$
holds with probability one in practice 
\cite{boshkovskaPracticalNonLinearEnergy2015,xiangRobustBeamformingWireless2012}.
This means we can construct $\mathbf{F}_{m,k}^{\star} = 
t^{\star} \mathbf{w}_{m} \mathbf{w}_{m}^H$ for  $t^{\star} > 0$
to unbound   objective   \eqref{eq:problem_eq2} when 
$t^{\star} \to +\infty$.
This contradicts the optimality of dual variables $ \left\{\lambda_m^{\star}\right\}$, 
$\left\{\gamma_{m,k}^{\star}\right\}$,  
$\left\{\mu_m^{\star}\right\}$, and $\left\{\iota_m^{\star}\right\}$.
Therefore, $\mathbf{B}_{m}^{\star} \succ \mathbf{0}$ with probability one.
If $\mathbf{B}_{m}^{\star} \succ \mathbf{0}$, then 
we must have $\mathrm{rank}\Big(\mathbf{S}_{m,k}^{\star}\Big) = N_{\mathrm{t}}-1 $.
}

If $\mathrm{rank}\Big(\mathbf{S}_{m,k}^{\star}\Big) = N_{\mathrm{t}}-1$, 
then we must have 
$\mathrm{rank}\Big(\mathbf{F}_{m,k}^{\star}\Big) = 1$ \cite{heFullDuplexCommunicationISAC2023a}.
Overall, we can conclude that the  optimal solutions
of problem \eqref{eq:problem_eq} are rank-one with probability one
when $\mathbf{a} \left(\theta_m\right) \notin \mathrm{span}
\Big( \bigcup_{i,j} \mathbf{h}_{m,i,j} \Big), \ \forall m$.

\section{Proof of Corollary \ref{corollary1}}
\label{app:corollary1}

The feasible set of problem \eqref{eq:problem_sensing_sca} is defined by the intersection of continuous constraints, which implies that it is closed. Furthermore, since the constraints ensure boundedness, the feasible set is compact.

The convex approximations in \eqref{sub1_1_21}, \eqref{sub1_1_1_21}, and \eqref{sub1_1_31} are derived using first-order Taylor expansions. These approximations serve as global lower bounds for the original nonconvex constraints while preserving their first-order properties. Furthermore, a feasible initial point is readily available provided that the original problem \eqref{eq:problem} is feasible.  {Consequently, the sequence of objective values generated by Algorithm \ref{alg:SCA} is guaranteed to converge according to \cite[Corollary 2.3]{beckSequentialParametricConvex2010}.}

{Because the generated sequence of iterates lies within a compact set, it admits at least one convergent subsequence.}
Assuming that Slater's condition holds at each iteration, we invoke \cite[Theorem 1]{marksGeneralInnerApproximationAlgorithm1978} to conclude that {the limit of
any convergent subsequence} generated by Algorithm~\ref{alg:SCA} is a KKT point.


\section{Proof of Lemma \ref{lemma3}}
\label{app:lemma2}
We   prove   Lemma \ref{lemma3} by contradiction. Let   
the optimal value of $\mathcal{P}\left(\eta,P,\epsilon\right)$
be $\tilde{a}$, then we must have $\mathcal{S}\left(\tilde{a} P,\epsilon\right) \geq \hat{\eta}$. 
Assume that $\mathcal{S}\left(\tilde{a} P,\epsilon\right) = e > \hat{\eta}$, 
then $\hat{\eta}$ is not the maximum value 
in the set $\left\{x | \mathcal{P}\left(x,P,\epsilon\right) = \mathcal{P}\left(\eta,P,\epsilon\right)\right\}$
because there  exists a value $e$ that is larger than $\hat{\eta}$,
which contradicts the definition of $\hat{\eta}$.
Therefore, we conclude $\mathcal{S}\left(\mathcal{P}\left( \eta,P,\epsilon \right) P,\epsilon\right) = \hat{\eta}$. 

Similarly, let the optimal value of $\mathcal{S}\left(P,\epsilon\right)$
be $\tilde{\eta}$ and the corresponding optimal solutions be $\left\{\mathbf{f}_{k}^{\star}\right\}$ and 
$q^{\star}$,
then we must have $\mathcal{P}\left(\tilde{\eta},P,\epsilon\right) \leq 1$.
Assume that $\mathcal{P}\left(\tilde{\eta},P,\epsilon\right) = e < 1$, 
then the solutions $\left\{\frac{1}{\sqrt{e}}  \mathbf{f}_{k}^{\star} \right\}$ and $\frac{q^{\star}}{e}$
are feasible for   
problem $\mathcal{S}\left(P,\epsilon\right)$
and achieve a higher objective value than $\tilde{\eta}$, which contradicts the fact 
that $\tilde{\eta}$ is the optimal value of $\mathcal{S}\left(P,\epsilon\right)$.
Therefore, we conclude $\mathcal{P}\left(\mathcal{S}\left(P,\epsilon\right),P,\epsilon\right) = 1$.

\section{Proof of Lemma  \ref{lemma4}}
\label{app:lemma3}
Given any semidefinite matrices $\left\{\mathbf{F}_{k} \right\}$ 
and a non-negative  scalar $q$ that satisfy the constraint in \eqref{SDPd232},
the  constraint in \eqref{2c_232} can always be satisfied 
by scaling  $\left\{\mathbf{F}_{k} \right\}$ and $q$ by a 
sufficiently large positive scalar.
However, such scaling does not necessarily guarantee that 
the constraint in \eqref{SDPc_232} is satisfied.
Therefore, the feasibility of problem \eqref{eq:problem232} 
is determined by the feasibility of constraint   \eqref{SDPc_232}.

We now demonstrate that 
if   $\mathbf{H}$ has full column rank, then
the constraint in \eqref{SDPc_232} can always be satisfied.
The condition that
$\mathbf{H}$ has full column rank implies
$\sum_{l=1}^{K-1}  
\mathbf{\Pi}_{\mathbf{e}^{l}_{k}}\mathbf{h}_k
\neq \mathbf{h}_k, \ \forall k$,
where $\left\{\mathbf{e}^{l}_{k}\right\}$ is a set of
orthogonal basis vectors  for $\mathrm{span}
\left( \mathbf{H}_k \right)$, 
and $\mathbf{H}_k$ is the submatrix of $\mathbf{H}$ obtained by removing its $k$-th column.
To satisfy the constraint, we
construct $\mathbf{F}_k= \alpha \mathbf{u}_k \mathbf{u}_k^H$ for each $k$,
where $\mathbf{u}_k = \left(\mathbf{I}_{N_{\mathrm{t}}}-\sum_{l=1}^{K-1}
    \mathbf{\Pi}_{\mathbf{e}^{l}_{k}}\right)\mathbf{h}_k, \ \forall k$.
Here, $\alpha$ is a positive scaling factor.
For each $k$, the equation $\mathbf{h}_j^H \mathbf{F}_k \mathbf{h}_j = 0, \ \forall j \neq k$ holds,
indicating that $\mathbf{F}_k$ introduces no interference to users $j \neq k$.
As a result, the constraint in \eqref{SDPc_232} reduces to
$\mathbf{h}_k^H \mathbf{F}_k \mathbf{h}_k   > \eta \sigma_{\mathrm{n}}^2, \ \forall k,$
which can  always be  satisfied by selecting a sufficiently large $\alpha$.

Therefore, if the condition that $\mathbf{H}$ has full column rank holds,
the problem in \eqref{eq:problem232} is always feasible.

\section{Proof of Corollary \ref{corollary3}}
\label{app:corollary3}
The proof of Corollary \ref{corollary3} is analogous to that of Corollary \ref{corollary1}. 
The primary technical challenge lies in verifying that the sequence generated by Algorithm~\ref{alg:SCA} is confined to a compact set. This requires establishing both the boundedness of the auxiliary variable $\varpi$ and the closedness of the constraint set in \eqref{2c_2}. First, the monotonicity of the objective value throughout the iterations implies that $\varpi$ is confined within a sublevel set bounded by its initial value $\varpi^{(0)} > 0$, thereby ensuring the boundedness of the sequence.

Next, we establish the closedness of the constraint set in \eqref{2c_2}. Let $\mathbf{M}(\mathbf{q}) = \boldsymbol{\Lambda} \hat{\mathbf{Z}} \boldsymbol{\Lambda}^T \succeq \mathbf{0}$ with $\mathbf{q} \succeq \mathbf{0}$, where
$\mathbf{q} = [q_1,  \ldots, q_M]^T$.
We define the extended-value function $f$ as
\begin{equation}
    f(\mathbf{q}) = 
    \begin{cases}
        \mathrm{tr}\big(\mathbf{M}^{-1}(\mathbf{q})\big), & \text{if } \mathbf{M}(\mathbf{q}) \succ \mathbf{0}, \\
        +\infty, & \text{otherwise}.
    \end{cases}
\end{equation}
The feasible set defined by \eqref{2c_2} is equivalent to the $\epsilon$-sublevel set of $f$, denoted by $\mathcal{S} = \{ \mathbf{q} \succeq \mathbf{0} \mid f(\mathbf{q}) \le \epsilon \}$. To prove that $\mathcal{S}$ is closed, consider a sequence $\{\mathbf{q}^{(r)}\} \subseteq \mathcal{S}$ converging to a limit point $\bar{\mathbf{q}}$. Since limit processes preserve non-strict inequalities, we have $\bar{\mathbf{q}} \succeq \mathbf{0}$. 
If $\mathbf{M}(\bar{\mathbf{q}})$ is singular, at least one eigenvalue of $\mathbf{M}(\mathbf{q}^{(r)})$ would tend to zero as $r \to \infty$. Consequently, $\lim_{r \to \infty} f(\mathbf{q}^{(r)}) = +\infty$, which contradicts the  condition $f(\mathbf{q}^{(r)}) \le \epsilon$ for all $r$. Therefore, $\mathbf{M}(\bar{\mathbf{q}})$ must be non-singular. By the continuity of the matrix inverse and the trace operator on the cone of positive definite matrices, we have $f(\bar{\mathbf{q}}) = \lim_{r \to \infty} f(\mathbf{q}^{(r)}) \le \epsilon$. Hence, $\bar{\mathbf{q}} \in \mathcal{S}$, confirming that the constraint set is closed.

Given that the iterative sequence is confined to a closed and bounded (i.e., compact) set, the convergence argument follows the same logic as the proof of Corollary \ref{corollary1}.

\bibliographystyle{IEEEtran}
\bibliography{ref.bib}

\end{document}